\documentstyle[12pt]{article}

\def\inf{\infty}
\newcommand{\be}{\begin{equation}}
\newcommand{\ee}{\end{equation}}
\newcommand{\ba}{\begin{array}{c}}
\newcommand{\ea}{\end{array}}
\newcommand{\bqa}{\begin{eqnarray}}
\newcommand{\eqa}{\end{eqnarray}}

\newcommand{\tr}{\hbox{tr}}
\newcommand{\IF}{{\cal J}}
\newcommand{\mathrm}[1]{\hbox{#1}}

\newcommand{\rlabel}{\label}
\newcommand{\rbibitem}{\bibitem}

\begin{document}

\begin{titlepage}
\begin{flushright}
CERN TH 6924/93\\
CPT-93 / P.2917\\
NORDITA-93/43 N,P\\
\end{flushright}
\vspace{2cm}
\begin{center}

\begin{Large}
\bf
LOW--ENERGY BEHAVIOR OF TWO--POINT FUNCTIONS\\
OF QUARK CURRENTS\\
\end{Large}

\vspace*{1cm}
        {\bf Johan Bijnens}\\
         NORDITA, Blegdamsvej 17\\
         DK--2100 Copenhagen \o, Denmark\\[1cm]
         {\bf Eduardo de Rafael}\\
         Centre de Physique Th\'eorique\\
         CNRS - Luminy, Case 907\\
         F 13288 Marseille Cedex 9, France\\[1cm]
         and \\[1cm]
         {\bf Hanqing Zheng}\\
         CERN, CH--1211 Geneva 23, Switzerland\\

\end{center}
\vfill
\begin{flushleft}
NORDITA-93/43 N,P\\
CPT-93 / P.2917\\
CERN-TH 6924/93\\
May   1993
\end{flushleft}
\end{titlepage}
\begin{titlepage}
\phantom{p}
\vfill
\begin{abstract}
We discuss vector, axial-vector,
 scalar and pseudoscalar two-point functions
at low and intermediate energies. We first review what is known from
chiral perturbation theory,
as well as from a heat kernel expansion within
the context of the extended Nambu-Jona-Lasinio (ENJL)
model of ref. \cite{12}. In this work
we derive then these two-point functions to all orders in the momenta and
to leading order in $1/N_c$. We find an improved high-energy behaviour
and a general way of parametrizing
them that shows relations between some of the two-point
functions, which are also valid in the presence of gluonic interactions.
The similarity between the shape
of the experimentally known spectral functions
and the ones we derive,
is greatly improved with respect to those predicted by
the usual constituent quark like models.
We also obtain the scalar mass
$M_S = 2 M_Q$ independent of the regularization scheme.
In the end, we calculate fully
an example of a nonleptonic matrix element in the ENJL--model,
the $\pi^+-\pi^0$ electromagnetic mass difference and find good agreement
with the measured value.
\end{abstract}
\vfill
\phantom{p}
\end{titlepage}

\section{INTRODUCTION AND REVIEW OF KNOWN RESULTS IN QCD}

\quad We shall be concerned with two--point functions of the vector,
axial--vector, scalar and pseudoscalar quark currents:
\begin{eqnarray}
\rlabel{1}
V^{(a)}_\mu (x)&\equiv& \bar q(x)\gamma_\mu{\lambda^{(a)}\over \sqrt{2}}
q(x)
\\
\rlabel{2}
A^{(a)}_\mu (x)&\equiv& \bar q(x)\gamma_\mu\gamma_5
{\lambda^{(a)}\over \sqrt{2}}
q(x)
\\
\rlabel{3}
S^{(a)} (x)&\equiv& -\bar q(x){\lambda^{(a)}\over \sqrt{2}}q(x)
\\
\rlabel{4}
P^{(a)} (x)&\equiv& \bar q(x)i\gamma_5 {\lambda^{(a)}\over \sqrt{2}}q(x)
\,
\end{eqnarray}
\noindent where $\lambda^{(a)}$ are Gell-Mann SU(3)--matrices acting on
the flavour triplets of Dirac spinors: $\bar q\equiv (\bar u(x),
\bar d(x), \bar s(x)).$ Summation over the colour degrees of freedom of
the quark fields is understood. These are the quark currents of the QCD
Lagrangian with three light flavours u, d, s, in the presence of external
vector $v_{\mu}(x)$, axial vector $a_\mu (x)$, scalar $s(x)$ and
pseudoscalar $p(x)$ field matrix sources; i.e.,

\be\rlabel{5}
{\cal L}_{QCD} (x)={\cal L}^0_{QCD} + \bar q\gamma_\mu (v_\mu +
\gamma_5 a_\mu)q - \bar q(s-i\gamma_5p)q\ ,
\ee
\noindent with

\be\rlabel{6}
{\cal L}^0_{QCD}=-{1\over 4}\sum_{A=1}^8 G_{\mu\nu}^{(A)} G^{(A)\mu\nu}
+ i\bar q\gamma^\mu (\partial_\mu +i G_\mu)q
\ee
\noindent and
\be\rlabel{7}
G_\mu\equiv g_s\sum_{A=1}^{N_C^2-1}{\lambda^{(A)}\over 2} G_\mu^{(A)}(x)
\ee
\noindent the gluon field matrix in the fundamental $SU(N_C=3)_{colour}$
representation, with $G^{(A)}_{\mu\nu}$ the gluon field strength tensor

\be\rlabel{8}
G^{(A)}_{\mu\nu}(x)=\partial_\mu G_\nu^{(A)}-\partial_\nu G^{(A)}_\mu
-g_sf_{ABC}G_{\mu}^{(B)}G_{\nu}^{(C)}\ ,
\ee
\noindent and $ g_s$
the colour coupling constant $(\alpha_s = g_s^2/4\pi)$.

We want to consider the set of two--point functions:

\bqa
\rlabel{9}
{ \Pi}^V_{\mu\nu} (q)_{ab}&=& i\int d^4x e^{iq\cdot x}
<0|T\left( V_\mu^{(a)} (x)V_\nu^{(b)}(0)\right) |0>
\\
\rlabel{10}
{ \Pi}^A_{\mu\nu} (q)_{ab}&=& i\int d^4x e^{iq\cdot x}
<0|T\left( A_\mu^{(a)} (x)A_\nu^{(b)}(0)\right) |0>
\\
\rlabel{11}
{ \Pi}^S_{\mu} (q)_{ab}&=& i\int d^4x e^{iq\cdot x}
<0|T\left( V_\mu^{(a)} (x)S^{(b)}(0)\right) |0>
\\
\rlabel{12}
{     \Pi}^P_{\mu} (q)_{ab}&=& i\int d^4x e^{iq\cdot x}
<0|T\left( A_\mu^{(a)} (x)P^{(b)}(0)\right) |0>
\\
\rlabel{13}
{     \Pi}^S (q)_{ab}&=& i\int d^4x e^{iq\cdot x}
<0|T\left( S^{(a)} (x)S^{(b)}(0)\right) |0>
\\
\rlabel{14}
{     \Pi}^P (q)_{ab}&=& i\int d^4x e^{iq\cdot x}
<0|T\left( P^{(a)} (x)P^{(b)}(0)\right) |0>\ .
\eqa

The relevance of these two--point functions to hadronic physics, within
the framework of current algebra, was emphasized a long time ago
(See refs.\cite{1} to \cite{4}.). With the advent of QCD it became possible
to compute the short--distance behavior of these two point functions
using perturbation theory in the colour coupling constant because of the
property of asymptotic freedom:
$\alpha_s(Q^2\equiv -q^2>> \Lambda^2_{QCD})\sim \log^{-1}
(Q^2/\Lambda^2_{QCD})$,
(See refs.\cite{5} to \cite{7}.). Non--perturbative
corrections at large $Q^2$
appear as inverse powers in $Q^2$ \cite{8}. The
inclusion of these corrections, combined with a phenomenological ansatz
 for the corresponding hadronic spectral functions at low energies, has
developped into the active
field of QCD sum rules. (See e.g. ref.\cite{9} for
a review and references therein.)
The QCD behaviour of the two point functions above at very low $Q^2$
values, is controlled by chiral perturbation theory ($\chi$PT). It is
important for our later purpose to review here what is known in this
respect.

{}From Lorentz--covariance and SU(3) invariance, the two--point functions
above can be decomposed in invariant functions as follows:

\be\rlabel{15}
{     \Pi^V_{\mu\nu} (q)_{ab}}= (q_\mu q_\nu -q^2g_{\mu\nu}){     \Pi_V^
{(1)}}(Q^2)\delta_{ab} + q_\mu q_\nu {     \Pi_V^{(0)}}(Q^2)\delta_{ab}
\ee

\be\rlabel{16}
{     \Pi^A_{\mu\nu} (q)_{ab}}= (q_\mu q_\nu -q^2g_{\mu\nu}){     \Pi_A^
{(1)}}(Q^2)\delta_{ab} + q_\mu q_\nu {     \Pi_A^{(0)}}(Q^2)\delta_{ab}
\ee

\bqa\rlabel{17}
{     \Pi^S_{\mu} (q)_{ab}}&=& {     \Pi_M^
{S}}(Q^2)q_\mu\delta_{ab}
\\
\rlabel{18}
{     \Pi^P_{\mu} (q)_{ab}}&=& {     \Pi_M^
{P}}(Q^2)iq_\mu\delta_{ab}
\\
\rlabel{19}
{     \Pi^S(q)_{ab}}&=& {     \Pi_
{S}}(Q^2)\delta_{ab}
\\
\rlabel{20}
{     \Pi^P(q)_{ab}}&=& {     \Pi_
{P}}(Q^2)\delta_{ab}\ .
\eqa

\noindent The
low energy behaviour of these invariant functions in $SU(2)_L\times
SU(2)_R$ $\chi$PT has
been worked out in ref.\cite{10}. It is easy to extend
their analysis to $SU(3)_L\times SU(3)_R$ $\chi$PT. In the chiral limit,
where the quark mass matrix ${\cal M}\rightarrow 0$; and with neglect
of chiral loops, which are non leading in the $1/N_C$--expansion
\cite{16},
the results are as follows:

\be\rlabel{21}
{    \Pi_V^{(1)}}(Q^2)= -4(2H_1 + L_{10}) + {\cal O}(Q^2)
\ee

\be\rlabel{22}
{    \Pi_V^{(0)}}(Q^2)= 0
\ee

\be\rlabel{23}
{\Pi_A^{(1)}}(Q^2)= {2f_0^2\over Q^2}-4(2H_1 - L_{10}) + {\cal O}(Q^2)
\ee

\be\rlabel{24}
{    \Pi_A^{(0)}}(Q^2)= 0
\ee

\be\rlabel{25}
{    \Pi_M^S    }(Q^2)= 0
\ee

\be\rlabel{26}
{    \Pi_M^P    }(Q^2)= {2B_0f_0^2\over Q^2}
\ee

\be\rlabel{27}
{    \Pi^S    }(Q^2)= 8B_0^2(2L_8 + H_2) + {\cal O}(Q^2)
\ee

\be\rlabel{28}
{\Pi}^P (Q^2)={2B_0^2f_0^2\over Q^2} + 8B_0^2(-2L_8+H_2) + {\cal O}
(Q^2)\ .
\ee
\noindent The functions ${    \Pi_A^{(1)}}$, ${    \Pi_M^P}$ and ${
\Pi^P}$ get their leading behaviour from the pseudoscalar Goldstone pole.
The residue of the pole is proportional to the pion decay constant $f_0$
$(f_0\simeq f_\pi=93.3 MeV)$.
The constant $B_0$ is related to the vacuum
expectation value

\be\rlabel{29}
<0|\bar qq|0>_{|q=u,d,s}=-f_0^2B_0\left( 1+{\cal O}({\cal M})\right)\ .
\ee
The constants $L_8$, $L_{10}$, $H_1$ and $H_2$ are coupling constants
of the $O(p^4)$ effective chiral Lagrangian in the notation of
Gasser and Leutwyler \cite{11}. The constants $L_8$ and $L_{10}$ are known
from the comparison between $\chi$PT and low energy hadron phenomenology.
At the scale of the $\rho$ mass:

\be\rlabel{30}
L_8=(0.9\pm 0.3)\times 10^{-3}
\ee
\noindent and

\be\rlabel{31}
L_{10}=(-5.5\pm 0.7)\times 10^{-3}\ .
\ee
\noindent The constant $H_1$ and $H_2$ correspond to couplings which
involve external source fields only and therefore cannot be extracted
from experiment unambiguously.

It has been recently shown \cite{12} that the extended Nambu Jona--Lasinio
model (ENJL--model) describes the values of the low energy parameters
rather well. In ref.\cite{12}, various relations between the parameters
of a low energy effective Lagrangian were obtained that were independant
of the input parameters and possible low energy gluonic corrections.
Among these was the relation

\be\rlabel{32}
f_V^2M_V^2-f_A^2M_A^2=f_\pi^2
\ee
\noindent between couplings and masses of the lowest vector and
axial--vector mesons and the pion--decay constant. This corresponds to
what is usually called the first Weinberg sum rule \cite{1}. The second
Weinberg sum rule, in the lowest resonance saturation form i.e.,
$f_V^2M_V^4=f_A^2M_A^4$, was, however, not satisfied exactly though the
deviation was numerically small. Part of the motivation which triggered
our interest in the present work has been to understand the origin of
this unsatisfactory result. As we shall show, the low energy expansion
method used in ref.\cite{12} is inappropriate to draw conclusions about the
intermediate $Q^2$--behaviour of two--point functions. The relevant
contributions can however be easily summed using a Feynman--diagram
expansion of the four fermion couplings in the ENJL--model without
introducing collective fields. Section 3 gives the details of
this summation method. The same results could be obtained of course
using the collective field method, provided though that all higher
orders in the $Q^2$--expansion were kept.

It is well known that the Weinberg sum rules play a crucial role in
the calculation of the electromagnetic $\pi^+$--$\pi^0$ mass difference
in the chiral limit.
In a previous calculation of this mass difference \cite{13} within the
framework of the effective action approach \cite{14}, it was shown that the
quality of the matching between long--distance and short--distance
contributions to the photon--loop integral is still rather poor when
the vector and axial--vector spectral functions are replaced by
constituent quark spectral functions alone. We shall (see also ref.\cite{15})
discuss this problem again in section 4, within the framework of
the ENJL--model and with the full $Q^2/M_Q^2$  dependence summed. This
calculation, which as we shall see is rather succesful, provides the
first non--trivial example of a genuine one loop calculation in the
ENJL--model. There are other applications one can now think of doing;
in particular calculations at the next to leading order in the
$1/N_C$--expansion. We plan to investigate this in the future.

Throughout this paper we shall work in the chiral limit. The inclusion
of corrections, whenever neccesary, due to nonzero current quark masses
can however be done with the present technology.

The rest of the paper is organized as follows.
In section 2 we give an overview of the ENJL--model and the two-point
functions in this model in the low-energy limit.
We also describe here the parametrization
usually used to go beyond the low-energy expansion in this model. Section
3 is the mainstay of this work. Here we derive the two-point functions
to all orders in the momenta.
We also discuss how gluonic corrections can be
included and present numerical results. In the next section we use these
two-point functions to calculate completely in
the ENJL--model a non-leptonic
matrix-element, the $\pi^+-\pi^0$ electromagnetic mass difference.
In the last section we present our main conclusions. In the appendix we
derive the underlying relations that allow for the same type of results
for the two-point functions,
as those found in ref. \cite{12} for the low-energy constants.

\section{TWO--POINT FUNCTIONS IN THE ENJL--MODEL}
\subsection{A brief review of the ENJL--model}

\quad
The scenario suggested in ref.\cite{12}, assumes that at intermediate energies
 below or of the order of the spontaneous chiral symmetry breaking scale
$\Lambda_\chi $, the Lagrangian of the ENJL--model is a good effective
realization of the standard QCD Lagrangian i.e.,
\be\rlabel{33}
{\cal L}_{QCD}\rightarrow {\cal L_{QCD}^{\Lambda_\chi}} +
{\cal L}_{NJL}^{S,P} +{\cal L}_{NJL}^{V,A}+{\cal O}({1\over
\Lambda _\chi^4})
\ee
\noindent with
\be\rlabel{34}
{\cal L}^{S,P}_{NJL}(x) =
{8\pi^2 G_S(\Lambda_{\chi}) \over N_c \Lambda_{\chi}^2}
\sum_{a,b}(\bar q_R^a q_L^b)(\bar q_L^b q_R^a)
\ee
\noindent and
\be\rlabel{35}
{\cal L}^{V,A}_{NJL} (x)= -
{8\pi^2 G_V(\Lambda_{\chi}) \over N_c \Lambda_{\chi}^2}
\sum_{a,b}\left[(\bar q_L^a \gamma^{\mu} q_L^b)
(\bar q_L^b \gamma_{\mu} q_L^a)
+ (L \to R) \right].
\ee
\noindent Here a and b are u, d, s flavour indices and colour summation
within each quark bilinear bracket is implicit;
$q_L={1\over 2}(1-\gamma _5)q(x)$ and $q_R={1\over 2}(1+\gamma _5)q(x)$.
The couplings $G_S$ and $G_V$ are dimensionless quantities. In principle
they are calculable functions of the ratio of the cut--off scale
 $\Lambda _\chi $  to the renormalization scale $\Lambda _{\overline{MS}}$.
In practice the calculation requires knowledge of the non-perturbative
behaviour of QCD and $G_S$ and $G_V$ will be taken as independent unknown
constants. In choosing the forms (\ref{34}) and (\ref{35})
of these four quark
operators we have only kept those couplings that are allowed by
the symmetries of the original QCD Lagrangian, and which are leading in
the $1/N_C$--expansion \cite{16}. With one inverse power of $N_C$ pulled--out,
both couplings
$G_S$ and $G_V$ are ${\cal O}(1)$ in the large $N_C$ limit.
The $\Lambda_\chi $  index in ${\cal L}^{\Lambda_\chi }_{QCD}$ means
that only the low frequency modes of the quark and gluon fields are to
be included.

 The basic assumption in considering the ENJL--model as a good effective
 Lagrangian of QCD is that at intermediate energies below or of the
 order of the spontaneous chiral symmetry breaking scale, the operators
${\cal L}_{NJL}^{S,P}$ and ${\cal L}_{NJL}^{V,A}$ are the leading
operators of higher dimension which, due to the growing of their
couplings $G_S$ and $G_V$ as the ultraviolet cut-off approaches its
critical value from above, become relevant or  marginal.%\cite{17}

As is well known in the Nambu Jona-Lasinio model \cite{18}, the ${\cal L}_{NJL}
^{S,P}$ operator, for values of $G_S>1$, is at the origin of the
the spontaneous chiral symmetry breaking. It is this operator which
generates a constituent chiral quark mass term ($U$  is a unitary
3$\times$3 matrix which collects the pseudoscalar Goldstone field modes):

\be\rlabel{36}
-M_Q(\bar q_L U^\dagger q_R + \bar q_R Uq_L)= -M_Q\bar QQ\ ,
\ee
\noindent like the one which appears in the Georgi--Manohar model \cite{19};
as well as in the effective action approach of ref.\cite{14}.

 As discussed in ref.\cite{12}, the ${\cal L}^{V,A}_{NJL} $ operator is at the
 origin of an effective axial coupling of the constituent quark fields
 $Q(x)$ with the Goldstone modes

\be\rlabel{37}
{i\over 2} g_A\bar Q \gamma ^\mu \gamma _5\xi _\mu Q\ ,
\ee
\noindent with

\be\rlabel{38}
g_A={1\over 1+4G_V\epsilon \Gamma (0,\epsilon )}\ ,
\ \ \ \epsilon \equiv M_Q^2/\Lambda _\chi ^2
\ee
\noindent and $\Gamma (0,\epsilon )$ the incomplete gamma function:
\be
 \Gamma (0,\epsilon )=\int^{\inf}_{\epsilon}  {dz\over z} e^{-z}\ .
 \ee
\noindent For $\epsilon $ small,
\be
\Gamma (0,\epsilon )= -\log \epsilon -\gamma _E + {\cal O}(\epsilon )\ ,
\ee
\noindent with $\gamma _E=0.5772\cdot\cdot\cdot$, Euler's constant.
We recall that in eq.~(\ref{37})
\be\rlabel{39}
\xi_\mu =
i\left\{\xi^\dagger\left[\partial_\mu -i(v_\mu +a_\mu )\right]\xi
-\xi\left[\partial_\mu -i(v_\mu -a_\mu )\right]\xi^\dagger\right\} \ ;
\ee
\noindent with
\be\rlabel{40}
U=\xi \xi
\ee
\noindent and
\be\rlabel{41}
Q= Q_L+Q_R
\ee
\be\rlabel{42a,b}
Q_L=\xi q_L\ ,\ \ \ \ \ \ Q_R=\xi^\dagger      q_R\ .
\ee

 The appearance of incomplete gamma functions is due to the proper time
regularization which is used in calculating the fermion determinant via
the Seeley--De Witt expansion. The type of integrals which appear are
($\epsilon =M_Q^2/\Lambda _\chi ^2$)
\begin{eqnarray}
\int^{\inf}_{1/ \Lambda ^2_\chi }{d\tau \over \tau }
 {1\over 16\pi ^2\tau ^2}\tau ^n e^{-\tau M^2_Q}&=&
{1\over 16\pi ^2}{1\over (M^2_Q)^{n-2}}  \int^{\inf}_\epsilon {dz\over z}
e^{-z}z^{n-2}
\nonumber\\
&=&{1\over 16\pi ^2}{1\over(M_Q^2)^{n-2}}
\Gamma (n-2,\epsilon )\ ;
\nonumber
\end{eqnarray}
\be\rlabel{43}
n=1,2,3,\cdot \cdot \cdot
\ee

The result for the axial coupling $g_A$ in eq.~(\ref{38})
is the result to
leading order  in the $1/N_C$--expansion. In terms of Feynman diagrams
it can be understood as the infinite sum of constituent quark bubbles
shown in Fig.1a, where the cross at the end represents the pion field.
These are the diagrams generated by the $G_V$ -- four fermion coupling
to leading order in the $1/ N_C$--expansion \cite{19b}.
The quark propagators in Fig.1a are constituent quark propagators,
solutions of the Schwinger--Dyson equation in the large $N_C$
approximation, which diagrammatically is represented in Fig.1b.

In its simplest version where one assumes furthermore that all the
relevant gluonic effects for low energy physics can be absorbed in the
couplings $G_S$ and $G_V$, the ENJL--model has only three free
parameters: $G_S$, $G_V$ and $\Lambda _\chi $. We find it useful to
specify these in terms of $M_Q$, $\Lambda _\chi $ and $g_A$ instead.
For this purpose one should remember that
($\epsilon =M_Q^2/\Lambda _\chi ^2$)
\be\rlabel{44}
G^{-1}_S=\epsilon \Gamma (-1,\epsilon )=e^{-\epsilon }-\epsilon
\Gamma (0,\epsilon ) \ .
\ee

\subsection{Low $Q^2$--behaviour of two--point functions in the
ENJL--model.}

\quad       As we discussed in the previous section,
the low--energy behaviour of two
point function quark currents, is governed by the constants $f_\pi ^2$,
$B_0$, $L_8$, $L_{10}$, $H_1$ and $H_2$. In the ENJL--model, these
constants have been calculated in ref.\cite{12} with the results
($\epsilon =M_Q^2/\Lambda _\chi ^2$):
\be\rlabel{45}
f_\pi ^2={N_C\over 16\pi ^2}4M_Q^2g_A\Gamma (0,\epsilon )
\ee
\be\rlabel{46}
2H_1+L_{10}=-{N_C\over 16\pi ^2}{1\over 3}\Gamma (0,\epsilon )
\ee
\be\rlabel{47}
2H_1-L_{10}=-{N_C\over 16\pi ^2}g_A^2
{1\over 3}\left[ \Gamma (0,\epsilon )
-\Gamma (1,\epsilon )\right]
\ee
\be\rlabel{48}
L_{8}={N_C\over 16\pi ^2}{1\over 16}g_A^2\left[ \Gamma (0,\epsilon )
-{2\over 3}\Gamma (1,\epsilon )\right]
\ee
\be\rlabel{49}
H_{2}={N_C\over 16\pi ^2}{1\over 8}g_A^2\left[
\left( 1-4{\Gamma (0,\epsilon )\over \Gamma (-1,\epsilon )}\right)
\Gamma (0,\epsilon )
+{2\over 3}\Gamma (1,\epsilon )\right] \ .
\ee
In fact, the calculations made in ref.\cite{12} suggest a possible improvement
of the invariant fuctions in
eqs.~(\ref{21}) to (\ref{28}) when the contribution from
the corresponding resonance
propagators is also taken into account, with the
results
\be\rlabel{50}
{     \Pi_V^{(1)}}(Q^2)=-4(2H_1+L_{10})-{2f_V^2Q^2\over M_V^2+Q^2}\ ,
\ee
\be\rlabel{51}
{    \Pi_A^{(1)}}(Q^2)={2f_\pi ^2\over Q^2}-4(2H_1-L_{10})-{2f_A^2Q^2
\over M_A^2+Q^2}\ ,
\ee
\noindent and
\be\rlabel{52}
{    \Pi^S}(Q^2)=8B_0^2\left\{ 2\tilde L_8+\tilde H_2 +
{2c_m^2\over Q^2+M_S^2}\right\} \ .
\ee
\noindent For the other invariant functions, the results are the same as
in eqs.~(\ref{21}) to (\ref{28})
with the parameter values given in eqs.~(\ref{45}) to (\ref{49}).
Several comments on these results are in order:

 i) Two relations which follow from the ENJL--model \cite{12} are
 \be\rlabel{53a,b}\rlabel{53}
 2H_1-L_{10}= -     f_A^2/2\qquad \mathrm{and}
\qquad 2H_1+L_{10}=-     f_V^2/2\ .
\ee
\noindent These relations are of the same type as the first Weinberg sum
rule relation in eq.~(\ref{32})
i.e., they are independent of the input
parameters and possible low energy gluonic corrections. As first shown
in ref.\cite{20}, they are crucial to ensure that the low energy effective
theory is compatible with the known short--distance properties of the
underlying theory -- QCD. It is reassuring that the ENJL--model indeed
respects these constraints.
Using these relations, we can also write ${    \Pi_V^{(1)}}$ and ${
\Pi_A^{(1)}}$ in the form
\be\rlabel{54}
{    \Pi_V^{(1)}}={2f_V^2M_V^2\over M_V^2+Q^2}\ ,
\ee
\be\rlabel{55}
{    \Pi_A^{(1)}}={2f_\pi ^2\over Q^2}+{2f_A^2M_A^2\over M_A^2+Q^2}\ ;
\ee
\noindent a form much more similar to the usual vector meson dominance
(VMD) phenomenological parametrizations found in the literature.

ii) The constant $c_m$ in eq.~(\ref{52})
denotes the coupling
 \be\rlabel{56}
 c_m tr\left( S(x)\left[ \xi ^\dagger \chi \xi ^\dagger +\xi \chi
 ^\dagger \xi \right] \right)
 \ee
 \noindent in the effective scalar Lagrangian. Here
\be\rlabel{57}
\chi =2B_0\left[ s(x)+ip(x)\right] \ ,
\ee
\noindent with $s(x)$ and $p(x)$
the external scalar and pseudoscalar matrix
field sources. As discussed in ref.\cite{12}, the couplings $L_8$ and $H_2$
receive contributions both from the quark loop -- which we denote
$\tilde L_8$  and $\tilde H_2$ -- and from the integration of scalar
fields -- which we denote   $L_8^S$ and $H_2^S$; i.e.,
  \be\rlabel{58a,b}
L_8=\tilde L_8 + L_8^S\qquad{\hbox{and}}\qquad H_2= \tilde H_2 +H_2^S\ .
\ee
\noindent In fact
\be\rlabel{59}
{2c_m^2\over M_S^2}=2L_8^S + H_2^S = 2H_2^S=\\
{N_C\over 16\pi ^2 } {1\over 4} g_A^2 {\Gamma (0,\epsilon )\over
\Gamma (-1,\epsilon )^2
}\left[ \Gamma (-1,\epsilon )-2\Gamma (0,\epsilon )
\right] ^2\ .
\ee

iii) Equations (\ref{50}), (\ref{51}) and (\ref{52})
imply a specific form of the
resummation to all orders in an expansion in powers of $Q^2$. As we shall
see, this is not however the correct form which follows from the exact
resummation of Feynman diagrams, to leading order in the
$1/N_C$--expansion.

 We shall finally give the expressions for the masses $M_S$, $M_V$ and
 $M_A$ which are obtained in the ENJL--model in ref.\cite{12}:
 \be\rlabel{60}
 M_S^2=4M_Q^2{1\over 1-{2\over 3}{\Gamma (1,\epsilon )\over
 \Gamma (0,\epsilon )}}\ ,
 \ee
 \be\rlabel{61}
 M_V^2={3\over 2}{\Lambda_\chi^2\over G_V}
 {1\over \Gamma (0,\epsilon )}=6M_Q^2{g_A\over 1-g_A}\ ,
 \ee
\be\rlabel{62}
M_A^2=6M_Q^2{1\over 1-g_A}{1\over 1-{\Gamma (1,\epsilon )\over
\Gamma (0,\epsilon )}}\ .
\ee

\section{FULL $Q^2$--DEPENDENT TWO--POINT FUNCTIONS IN THE ENJL--MODEL}

\subsection{The Vector Two--Point Function}
\quad To illustrate
the method, we shall first discuss with quite a lot of
detail the vector invariant function ${    \Pi}_V^{(1)}(Q^2)$. In the
ENJL--model, and to leading order in the $1/N_C$--expansion, we have to
sum over the infinite class of bubble diagrams shown in Fig. 2a.
Algebraically, this corresponds to the sum
\be\rlabel{63}
(q^\mu q^\nu -q^2g^{\mu \nu })\overline{    \Pi}_V^{(1)}
+(q^\mu q^\alpha -q^2g^{\mu \alpha })\overline{    \Pi}
_V^{(1)}\left( {-4\pi^2G_V\over N_C\Lambda _\chi ^2}\right)
\times 2(q_\alpha q^\nu -q^2g_\alpha ^\nu )
    {\overline\Pi}_V^{(1)}+\cdot\cdot\cdot \ ,
\ee
\noindent where the explicit factor of 2 in the second term comes from
the two possible contractions between the fermion fields of the vector
four--quark operator. The overall result is
\be\rlabel{64}
(q^\mu q^\nu -q^2g^{\mu \nu })\left\{      {\overline\Pi}_V^{(1)}
+    {\overline\Pi}_V^{(1)}
q^2{8\pi ^2G_V\over N_C\Lambda _\chi ^2}    {\overline\Pi}_V^{(1)}
+\cdot\cdot\cdot\right\}\ .
\ee
\noindent Notice that for this two-point function, only the vector four
quark interaction with coupling $G_V$ can contribute. No mixing between
different operators can occur in this case. The one loop bubble in
Fig.2b corresponds to the bare fermion--loop diagram of the mean field
approximation defined in ref.\cite{12}, which in what follows we shall
denote with an overlined expression
\be\rlabel{65}
 {^1\Pi}_V^{(1)}(Q^2) =    {\overline\Pi}_V^{(1)}(Q^2)\ .
\ee
\noindent It is easy to see that at the n--loop bubble level, the
corresponding  expression for ${^n\Pi}_V^{(1)}(Q^2)$ will be
given by the $n-1$ -- loop bubble result multiplied by the coupling
$G_V$ and one more factor of the one-loop result,
i.e.,
\be\rlabel{66}
{^n{\Pi}}_V^{(1)}(Q^2)= {^{n-1}\Pi}_V^{(1)}(Q^2){8\pi^2G_V\over
N_C\Lambda _\chi ^2}(-Q^2){\overline\Pi}_V^{(1)}(Q^2)\ .
\ee
\noindent This series can then be summed with the result
\be\rlabel{67}
{\Pi}_V^{(1)} =    {{\overline\Pi}_V^{(1)}(Q^2)\over 1+Q^2 {8\pi^2G_V
\over N_C\Lambda _\chi ^2}     {\overline\Pi}_V^{(1)}(Q^2)}\ .
\ee

We discuss next the calculation of
$    {\overline\Pi}_V^{(1)}(Q^2)$ in some
detail. The spectral function associated to $    {\overline\Pi}_V^{(1)}(Q^2)$
is the one corresponding to the $Q\bar Q$ intermediate state in a P--wave
and can be
calculated unambiguously  with the well known result
\be\rlabel{68}
{1\over \pi}Im    {\overline\Pi}_V^{(1)}(t)={N_C\over 16\pi ^2}
{4\over 3}\left( 1+{2M_Q^2\over t}\right)
\sqrt{1-{4M_Q^2\over t}} \theta (t-4M_Q^2)\ .
\ee
\noindent The function $    {\overline\Pi}_V^{(1)}(Q^2)$ we seek for has to
obey three criteria:

i) It must obey the relevant Ward identities.

ii) Its discontinuity should coincide with the spectral function in
eq.~(\ref{68}).

iii) When expanded in powers of $Q^2$ it must reproduce the heat kernel
calculation of the effective action approach, with the same proper time
regularization results.

\noindent These three criteria will in fact apply to all the two--point
functions we discuss.

To proceed  with the calculation of $    {\overline\Pi}_V^{(1)}(Q^2)$, we
 write a once subtracted dispersion relation for this function
\be\rlabel{69}
    {\overline\Pi}_V^{(1)}(Q^2)=    {\overline\Pi}_V^{(1)}(0)-Q^2\int^{\inf}_0
 {dt\over t}{1\over t+Q^2}{1\over \pi }Im    {\overline\Pi}_V^{(1)}(t)\ .
\ee
\noindent Requirement iii) implies that $    {\overline\Pi}_V^{(1)}(0)$ is
 fixed, with the result ($\epsilon =M_Q^2/\Lambda _\chi ^2$)
 \be\rlabel{70}
    {\overline\Pi}_V^{(1)}(0)={N_C\over 16\pi ^2}{4\over 3}
\Gamma (0,\epsilon )\ .
\ee
\noindent With ${1\over \pi }Im    {\overline\Pi}_V^{(1)}(t)$ in
eq.~(\ref{68})
inserted in the integrand of the r.h.s. of eq.~(\ref{69});
and with the
 successive change of variables
\be\rlabel{72a,b}
{4M_Q^2\over t}=1-y^2 \,\,\,{\hbox{and}} \,\,\, y=1-2x\ ,
\ee
\noindent we have that
\begin{displaymath}
\int^{\inf}_0{dt\over t}{Q^2\over t+Q^2}{1\over \pi }
Im    {\overline\Pi}_V^{(1)}(t)=
\end{displaymath}
\be\rlabel{73}
{N_C\over 16\pi ^2}{2\over 3}
{\int^{1}_0}dx(1-2x)^2\left[ 1+2x(1-x)\right]
{Q^2\over M_Q^2+Q^2x(1-x)}\ .
\ee
\noindent In order to match with the proper time regularization which
has been used in the calculation of $    {\overline\Pi}_V^{(1)}(0)$, we next
replace the denominator in the r.h.s. of eq.~(\ref{73}) as follows
\be\rlabel{74}
{1\over M_Q^2+Q^2x(1-x)}\to \int^{\inf}_{1/\Lambda _\chi ^2}d\tau
e^{-\tau [M_Q^2+Q^2x(1-x)]}\ .
\ee
\noindent Performing  an integration by parts in the $x$--variable we
 then finally get the result ($\epsilon =M_Q^2/\Lambda _\chi ^2$)
\be\rlabel{75}
\int^{\inf}_0{dt\over t}{Q^2\over t+Q^2}{1\over \pi }
Im    {\overline\Pi}_V^{(1)}(t)= {N_C\over 16\pi ^2}{4\over 3}\left\{
\Gamma (0,\epsilon )-6\int^1_0dx x(1-x)\Gamma (0,x_Q)\right\}
\ee
\noindent where $x_Q$ is a short--hand notation, which we shall use
from here onwards, for
\be\rlabel{76}
x_Q={M_Q^2+Q^2x(1-x)\over \Lambda _\chi ^2}\ .
\ee
\noindent Combining eqs.~(\ref{69}), (\ref{70}) and (\ref{75}), we obtain
\be\rlabel{77}
    {\overline\Pi}_V^{(1)}(Q^2)={N_C\over 16\pi ^2}{8}\int^1_0dx x(1-x)
\Gamma (0,x_Q )\ .
\ee
\noindent The  first few terms in a $Q^2$--expansion of this expression
are
\be\rlabel{78}
    {\overline\Pi}_V^{(1)}(Q^2)={N_C\over 16\pi ^2} \left\{ {4\over 3}
\Gamma (0,\epsilon )-{4\over 15}\Gamma (1,\epsilon ){Q^2\over M_Q^2}
+{1\over 35} \Gamma (2,\epsilon ) {Q^4\over M_Q^4} +
 {\cal O}(Q^6)\right\}\ ,
\ee
\noindent in agreement with the proper time regularized heat kernel
effective action result.

The imaginary part of ${\overline\Pi}_V^{(1)}(Q^2)$ evaluated from
eq.~(\ref{77})
using the $i\epsilon $ prescription $x_Q\to \left( M_Q^2 +
Q^2x(1-x)-i\epsilon \right) /\Lambda _\chi ^2$ in the log term: $\Gamma
(0,x_Q) = -\log x_Q-\gamma _E+{\cal O}(x_Q)$, reproduces the spectral
function in eq.~(\ref{68}).

We shall now try to cast the result for ${\Pi}_V^{(1)}(Q^2)$ in
eq.~(\ref{67}) in the simple VMD--form of eq.(\ref{54}):
\be\rlabel{79}
{    \Pi}_V^{(1)}(Q^2)=
{N_C\Lambda _\chi ^2/8\pi ^2G_V\over {({\overline\Pi}_V^{(1)})}^{-1}
{N_C\Lambda _\chi ^2\over  8\pi ^2G_V}  + Q^2} =
{2f_V^2(Q^2)M_V^2(Q^2)\over  M_V^2(Q^2)+Q^2}\ ,
\ee
\noindent where we have set
\be\rlabel{80}
  2f_V^2(Q^2)M_V^2(Q^2)= {N_C\Lambda _\chi ^2\over 8\pi ^2G_V}\ ,
\ee
\noindent and
\be\rlabel{81}
M_V^2(Q^2)={\Lambda _\chi ^2\over 4G_V}
 {1\over \int^1_0 dx x(1-x)\Gamma (0,x_Q)}\ .
\ee
\noindent We find that the full $Q^2$--dependent vector two--point
function can indeed be cast in the VMD--form of eq.~(\ref{54})
 provided
that the coupling parameters $f_V(Q^2)$ and $M_V(Q^2)$ become $Q^2$
dependent. Their value at $Q^2=0$ happen to coincide, in this case,
with the couplings in the low energy effective Lagrangian i.e.,
\be\rlabel{82}
 M_V^2(Q^2=0)=M_V^2={3\over 2} {\Lambda _\chi ^2\over G_V}
 {1\over \Gamma (0,\epsilon )} \ .
\ee
\noindent and
\be\rlabel{83}
f_V^2(Q^2=0) = f_V^2 = {N_C\over 16\pi ^2} {2\over 3}
 \Gamma (0,\epsilon )  \ .
\ee
\noindent The product $f_V^2(Q^2)M_V^2(Q^2)$ is scale--invariant.

In order to see the hadronic content of the full vector  two--point
function we propose to examine the complete spectral function
 \be\rlabel{84}
{1\over \pi } Im{    \Pi}_V^{(1)}(t) = {{1\over \pi }
Im    {\overline\Pi}_V^{(1)}(t)\over
{\left[ 1-t{8\pi ^2G_V\over N_C\Lambda _\chi ^2}
Re    {\overline\Pi}_V^{(1)}(t)\right] }^2 +{\left[ t
{8\pi ^2G_V\over N_C\Lambda _\chi ^2}
Im    {\overline\Pi}_V^{(1)}(t)\right] }^2 }
\ee
and plot it as a function of t for the input parameter values
\be\rlabel{85a,b}
M_Q=265 MeV,\ \ \Lambda _\chi  = 1165 MeV \ee
\noindent and
\be\rlabel{85}
g_A=0.61  \ .
\ee
 \noindent These are the values corresponding to fit $\#1$ in ref.\cite{12}.
The plot is the one shown in Fig.~3 (the full line). For the sake
of comparison, we have also ploted in the same figure the spectral
function ${1\over \pi }Im    {\overline\Pi}_V^{(1)}(t)$  corresponding to
 the mean field approximation (the dashed line). The improvement towards
a reasonable simulation of the well known experimental shape of the
$J^P=1^-,\ I=1$ hadronic spectral function is rather notorious.

\subsection{The Axial--Vector Two--Point Function}

\quad The
infinite series of bubble diagrams we have to sum in this case is
formally  very similar to the one already discussed in the previous
subsection. Again, only the vector four--quark interaction with coupling
$G_V$ contributes in this case with the result
\be\rlabel{86}
{    \Pi}_A^{(1)}(Q^2) =
{      {\overline\Pi}_A^{(1)}(Q^2) \over 1 + Q^2{8\pi ^2G_V\over
N_C\Lambda _\chi ^2}    {\overline\Pi}_A^{(1)}(Q^2) }\ .
\ee

 The axial two--point function $    {\overline\Pi}_A^{\mu \nu }(q)$ in the
mean field approximation, has however more structure than the
corresponding vector two--point function because now,
due to the presence of a constituent quark mass, the axial invariant
function $    {\overline\Pi}_A^{(0)}$ in the decomposition corresponding to
eq.~(\ref{16})
doesn't vanish. Both
${    {\overline\Pi}}_A^{(1)}(Q^2)$
and ${    {\overline\Pi}}_A^{(0)}(Q^2)$ have associated spectral functions
which can be calculated unambiguously with the results:
\be\rlabel{87}
{1\over \pi }Im
    {\overline\Pi}_A^{(1)}(t)={N_C\over 16\pi ^2}{4\over 3} \left(
1-{4M_Q^2\over t } \right) \sqrt{1-{4M_Q^2\over t}} \theta (t-4M_Q^2)
\ee
\noindent
\be\rlabel{88}
{1\over \pi }Im
    {\overline\Pi}_A^{(0)}(t)={N_C\over 16\pi ^2}
{8M_Q^2\over t }  \sqrt{1-{4M_Q^2\over t}} \theta (t-4M_Q^2)
\ee
Notice that the three spectral functions
${1\over \pi}Im    {\overline\Pi}_V^{(1)}(t)$,
${1\over \pi}Im    {\overline\Pi}_A^{(1)}(t)$
and ${1\over \pi}Im    {\overline\Pi}_A^{(0)}(t)$
satisfy the identity:
\be\rlabel{89}
{1\over \pi}Im    {\overline\Pi}_V^{(1)}(t)-
{1\over \pi}Im    {\overline\Pi}_A^{(1)}(t)-
{1\over \pi}Im    {\overline\Pi}_A^{(0)}(t)=0\ .
\ee
\noindent In fact this is nothing but a particular case of a general
Ward identity which two--point functions in the mean field approximation
must obey:
\be\rlabel{90}
    {\overline\Pi}_V^{(1)}(Q^2)-
    {\overline\Pi}_A^{(1)}(Q^2)-
    {\overline\Pi}_A^{(0)}(Q^2)=0\ .
\ee
\noindent The proof of this identity can be found in the Appendix.
It is precisely this identity which guarantees that the first Weinberg
sum rule in the mean field approximation is automatically satisfied.

{}From the asympotic behaviour of
$Im    {\overline\Pi}_A^{(0)}(t)$ in eq.~(\ref{88})
we conclude that the dispersive
part of the function
$    {\overline\Pi}_A^{(0)}(Q^2)$ obeys an unsubtracted dispersion relation.
To this we have to add the pole term calculated in the effective action
 approach i.e.,
 \be\rlabel{91}
    {\overline\Pi}_A^{(0)}(Q^2)=
{-2\bar f_\pi ^2\over Q^2} + \int^{\inf}_0   {dt\over t+Q^2}{1\over \pi }
Im    {\overline\Pi}_A^{(0)}(t) \ ,
\ee
\noindent with ($\epsilon =M_Q^2/\Lambda _\chi ^2$)
\be\rlabel{92}
 {\bar f}_\pi ^2
 ={N_C\over 16\pi ^2}4M_Q^2\Gamma (0,\epsilon )\ .
\ee
\noindent Using the same change of variables as in
eqs.~(\ref{72a,b}), we obtain
\be\rlabel{93}
\int^{\inf}_0{dt\over t+Q^2}{1\over \pi }Im    {\overline\Pi}_A^{(0)}(t)
={N_C\over 16\pi ^2}4M_Q^2 \int^{1}_0
dx(1-2x)^2 {1\over M_Q^2+Q^2x(1-x)}\ .
\ee
\noindent Next, we use the same proper time representation for the
denominator in the right hand side as in eq.~(\ref{74}),
and perform an
integration by parts in the $x$--variable,  with the result
\be\rlabel{94}
\int^{\inf}_0{dt\over t+Q^2}{1\over \pi }Im    {\overline\Pi}_A^{(0)}(t)
={N_C\over 16\pi ^2}{8M_Q^2\over Q^2}
\left[ \Gamma (0,\epsilon )-\int^{1}_0dx\Gamma (0,x_Q)\right] \ ,
\ee
\noindent with $x_Q$ defined in eq.~(\ref{76}).
Inserting this result in
the r.h.s. of eq.~(\ref{91}) leads to the final result
\be\rlabel{95}
     {\overline\Pi}_A^{(0)}(Q^2)=-{N_C\over 16\pi ^2}{8M_Q^2\over Q^2}
\int^{1}_0dx\Gamma (0,x_Q)\equiv -{2\bar f_\pi ^2(Q^2)\over Q^2}\ ,
\ee
\noindent which defines a running $\bar f_\pi(Q^2) $  in the mean field
approximation.
The first few terms in a $Q^2$--expansion of $\bar f_\pi ^2(Q^2)$ are
\be\rlabel{96}
 \bar f_\pi ^2(Q^2)={N_C\over 16\pi ^2}4M_Q^2
 \left\{ \Gamma (0,\epsilon )-{1\over 6}\Gamma (1,\epsilon )
 {Q^2\over M_Q^2}+{\cal O}(Q^4)\right\} \ ,
\ee
\noindent in agreement with the proper time regularized heat kernel
effective action result.

Once we have calculated ${    \overline\Pi}_A^{(0)}(Q^2)$ and
$    {\overline\Pi}_V^{(1)}(Q^2)$, the function
${\overline\Pi}_A^{(1)}(Q^2)$ follows from the Ward identity in eq.~(\ref{90})
with the result
\be\rlabel{97}
    {\overline\Pi}_A^{(1)}(Q^2)= {N_C\over 16\pi ^2} 8\left\{ {M_Q^2\over Q^2}
\int^{1}_0dx\Gamma (0,x_Q)+\int^1_0dx x(1-x)\Gamma (0,x_Q)\right\}\ .
\ee
\noindent It is nevertheless instructive to calculate
$    {\overline\Pi}_A^{(1)}(Q^2)$
independently as an illustration of the method we are using.
First we observe that the dispersive part of $    {\overline\Pi}_A^{(1)}(Q^2)$
needs a subtraction. On the other hand, we know from the effective action
calculation that $    {\overline\Pi}_A^{(1)}(Q^2)$  has a pole term (see
eqs.~(\ref{51}) and (\ref{47}))
\be\rlabel{98}
    {\overline\Pi}_A^{(1)}(Q^2)
= {2\bar f_\pi ^2\over Q^2} -{N_C\over 16\pi ^2} {4\over 3}
\Gamma (1,\epsilon )   + {N_C\over 16\pi ^2}{4\over 3}
\Gamma (0,\epsilon )  +{\cal O}(Q^2)\ .
\ee
\noindent We recognize the second term in the r.h.s. of this expression
as the constant term in the $Q^2$--expansion of $\bar f_\pi ^2(Q^2)$
in eq.~(\ref{96}),
which means that the subtraction constant needed for the
 dispersion relation is only the third term. However this term is
 precisely the same as the one corresponding to the vector function
${\Pi}_V^{(1)}(0)$ in eq.~(\ref{70}).
We must therefore separate the
spectral function ${1\over \pi }Im{    \Pi}_A^{(1)}(t)$ in two pieces:
one which reproduces the vector spectral function ${1\over \pi } Im
{\Pi}_V^{(1)}(t)$, for which we shall write a once subtracted
dispersion and the rest. But this is precisely the Ward identity
separation we already pointed out in eq.~(\ref{89}).
We then have
\begin{eqnarray}
{    \Pi}_A^{(1)}(Q^2)&=& {2\bar f_\pi ^2\over Q^2} -{N_C\over 16\pi ^2}
\int^{\inf}_0 {dt\over t+Q^2}{8M_Q^2\over t} \sqrt{1-{4M_Q^2\over t}}
\theta (t-4M_Q^2) +
\rlabel{99}
 {N_C\over 16\pi ^2}{4\over 3}\Gamma (0,\epsilon )\nonumber\\&&
-{N_C\over 16\pi ^2}{4\over 3} \int^{\inf}_0{dt\over t}{Q^2\over t+Q^2}
\left( 1+ {2M_Q^2\over t}\right) \sqrt
{1-{4M_Q^2\over t}} \theta(t-4M_Q^2)
\ .\end{eqnarray}
\noindent the two dispersive integrals are those calculated before in
eq.~(\ref{94})
for the unsubtracted piece. Putting these results together
leads to the result in eq.~(\ref{97}).

  With these results in hand, we shall now try to cast
${\Pi}_A^{(1)}(Q^2)$ in eq.~(\ref{86}) as close as possible to the
VMD--form of eq.~(\ref{55}):
\be\rlabel{100}
    {    \Pi}_A^{(1)}(Q^2) ={
    {\overline\Pi}_V^{(1)}-
    {\overline\Pi}_A^{(0)} \over 1-Q^2{8\pi ^2G_V\over N_C\Lambda _\chi ^2}
    {\overline\Pi}_A^{(0)}+ Q^2{8\pi ^2G_V\over N_C\Lambda _\chi ^2}
    {\overline\Pi}_V^{(1)} }\ .
\ee
\noindent From the calculated expression for
${\overline\Pi}_A^{(0)}(Q^2)$  in eq.~(\ref{95}), it follows that
\be\rlabel{101}
1-Q^2{8\pi ^2G_V\over N_C\Lambda _\chi ^2}
    {\overline\Pi}_A^{(0)}(Q^2) =1 + 4G_V {M_Q^2\over \Lambda _\chi ^2}
\int^1_0dx\Gamma (0,x_Q)\ .
\ee
\noindent At $Q^2=0$, the r.h.s. is precisely $g_A^{-1}$
(see eq.~(\ref{38})).
This is not a surprising result. The evaluation of the axial vector form
factor of a constituent chiral quark from
the infinite series of bubble graphs in Fig.~1a, leads to the result
 \be\rlabel{102}
 g_A(Q^2)={1\over  1+ {(G_V/\Lambda_\chi^2)}
  4M_Q^2\int^1_0 dx \Gamma (0,x_Q)} \ ,
\ee
\noindent which at $Q^2=0$ coincides with the axial coupling constant
 $g_A$ obtained in the calculation of the low energy effective action
 in ref.\cite{12}. Using this result, we can now rewrite the r.h.s. of
eq.~(\ref{100}) in the following simple form:
 \be\rlabel{103}
 {    \Pi}_A^{(1)}(Q^2)
 =  {2f_\pi^2(Q^2)\over Q^2}
 + {2f_A^2(Q^2)M_A^2(Q^2)\over M_A^2(Q^2)+Q^2}\ ,
 \ee
\noindent where
\be\rlabel{104}
f_\pi ^2(Q^2) = g_A(Q^2)\bar f_\pi ^2(Q^2) \ ;
\ee
\be\rlabel{105}
M_A^2(Q^2)={1\over g_A(Q^2)} M_V^2(Q^2)\ ;
\ee
\noindent and
\be\rlabel{106}
f_A^2(Q^2)=g_A^2(Q^2)f_V^2(Q^2)\ ,
\ee
\noindent with $\bar f_\pi ^2(Q^2)$, $M_V^2(Q^2)$ and $f_V^2(Q^2)$
 as given in eqs.~(\ref{95}), (\ref{81}) and (\ref{80}), respectively.
    Notice that at $Q^2=0$
\be\rlabel{107}
 f_\pi ^2(Q^2=0) = f_\pi ^2 = {N_C\over 16\pi ^2} 4M_Q^2 g_A
    \Gamma (0,\epsilon ) \ ;
 \ee
\noindent i.e., the same expression  which appears in the low energy
effective action; however $M_A^2(Q^2=0)$ and $f_A^2(Q^2=0)$
do not coincide with the expressions for the couplings $M_A^2$ and $f_A^2
$  as given in eqs.~(\ref{62}) and (\ref{53}), (\ref{47}).
We now have instead
\be\rlabel{106a,b}
M_A^2(Q^2=0)={1\over g_A} M_V^2 \,\,\,{\hbox{and}} \,\,\,
 f_A^2(Q^2=0)=g_A^2f_V^
2\ .\ee
The remarkable new result is that now, in terms of the running couplings
and running masses, both the first and second Weinberg sum rule are
satisfied
\be\rlabel{109}
f_V^2(Q^2)M_V^2(Q^2)=f_A^2(Q^2)M_A^2(Q^2)+f_\pi ^2(Q^2)
\ee
\noindent and
\be\rlabel{110}
f_V^2(Q^2)M_V^4(Q^2)=f_A^2(Q^2)M_A^4(Q^2)\ .
\ee

The fact that $f_A^2(Q^2=0)$ does not coincide with the $f_A^2$ defined
 in the effective action approach can be understood from a comparison
between eqs.~(\ref{55}) and eqs.~(\ref{103}).
Both expressions coincide to ${\cal O}(Q^2)$:
\be\rlabel{111a}
{\Pi}_A^{(1)}|_{eq.~(\ref{55})}
= {2f_\pi ^2\over Q^2} + 2f_A^2 +{\cal O}(Q^2
)\ee
\be\rlabel{111b}
{\Pi}_A^{(1)}|_{eq.~(\ref{103})}
= {2f_\pi ^2\over Q^2} -{N_C\over 16\pi ^2}
{4\over 3} g_A^2 \Gamma (1,\epsilon )
+ 2f_A^2(Q^2=0) +{\cal O}(Q^2)
\ee
\noindent It is the fact that part of the constant term is reabsorbed
in the $Q^2$--dependence of $f_\pi ^2$, that is at the origin of this
difference. The precise relation is
\be\rlabel{112}
f_A^2=f_A^2(Q^2=0)- {N_C\over 16\pi ^2} {4\over 3}g_A^2
\Gamma (1,\epsilon )
\ee
\subsection{The Scalar Two--Point Function}
  The full scalar invariant function from the sum of the infinite
series of bubble diagrams has the form
\be\rlabel{113}
  {    \Pi}_S(Q^2)={ {\overline\Pi}_S(Q^2) \over
  1- {4\pi ^2G_S\over N_C\Lambda _\chi ^2}{\overline\Pi}_S(Q^2) }\ .
\ee
\noindent It only involves the four--quark operator with $G_S$--coupling.
As in the previous vector and axial--vector discussion, we now proceed to
the calculation of the scalar two--point function
 ${    \overline\Pi}_S(Q^2)$ in the mean field approximation. The associated
spectral function can be calculated unambiguously, with the result
\be\rlabel{114}
 {1\over \pi } Im     {\overline\Pi}_S(t)= {N_C\over 16\pi ^2} 8M_Q^2
 \left( {t\over 4M_Q^2}-1\right) \sqrt{
 1-{4M_Q^2\over t}}\theta (t-4M_Q^2)
\ .\ee
\noindent From its assymptotic behaviour at large t, it follows that
the function ${\overline\Pi}_S(Q^2)$ obeys a dispersion relation with
two--subtractions for the term proportional to $t\sqrt{1-4M_Q^2/t}$ and
one subtraction for the term $M_Q^2\sqrt{1-4M_Q^2/t}$. Accordingly, we
write the dispersion relation
\begin{displaymath}
{\overline\Pi}_S(Q^2)={\overline\Pi}_S(0)+Q^2\left( {\overline\Pi}_S'(0)
-{N_C\over 16\pi ^2}{4\over 3} \Gamma(1,\epsilon)\right)
\end{displaymath}
\begin{displaymath}
-Q^2\int^{\inf}_0{dt\over t} {1\over t+Q^2}{N_C\over 16\pi ^2} (-8M_Q^2)
\sqrt {1-{4M_Q^2\over t}}\theta (t-4M_Q^2)
\end{displaymath}
\be\rlabel{115}
+Q^4\int^\inf_0{dt\over t^2} {1\over t+Q^2} {N_C\over 16\pi ^2} (2t)
\sqrt {1-{4M_Q^2\over t}}\theta (t-4M_Q^2)\ ;
\ee
\noindent where
${\overline\Pi}_S(0)$ and ${ \overline\Pi}_S'(0)$ are
already known from the effective action calculation i.e.,
\be\rlabel{116}
{    \overline\Pi}_S(0)= {N_C\over 16\pi ^2} 4M_Q^2
\left[ \Gamma (-1,\epsilon )-2\Gamma (0,\epsilon )\right] \ ;
\ee
\noindent and
\be\rlabel{117}
{    \overline\Pi}_S'(0)= -{N_C\over 16\pi ^2} 2\left(
\Gamma (0,\epsilon ) -{2\over 3}\Gamma (1,\epsilon)\right) \ .
\ee
\noindent Notice that only the divergent pieces (i.e., terms proportional
to $\Gamma (-1,\epsilon )$ and $\Gamma (0,\epsilon )$;
but not $\Gamma (n,\epsilon )$, $n\simeq 1$) are
retained in the subtraction constant.
The integral we have to compute is
\be\rlabel{118}
\int^{\inf}_{4M_Q^2}
{dt\over t} {1\over  t+Q^2} \sqrt {1-{4M_Q^2\over t}}
={1\over 2} \int^1_0 dx (1-2x)^2{1\over  M_Q^2 + Q^2x(1-x)}\ ,
\ee
\noindent where we have made the standard change of variables of
eq.~(\ref{72a,b}). Using the representation of eq.~(\ref{74}),
and doing one
integration by parts in the variable $x$,
and replacing  eqs.~(\ref{116}) and (\ref{117}),
we obtain the result
 ($\epsilon ={M_Q^2\over \Lambda _\chi ^2}$, $x_Q={M_Q^2+Q^2x(1-x)\over
 \Lambda _\chi ^2}$ )
\be\rlabel{120}
{    \overline\Pi}_S(Q^2)=
{N_C\over 16\pi ^2} 4M_Q^2\left\{ \Gamma (-1,\epsilon )
-2\left( {Q^2\over 4M_Q^2}+1\right)  \int^1_0 dx \Gamma (0,x_Q)\right\}
\ .\ee

We shall next evaluate the full ${\Pi}_S(Q^2)$ function in
eq.~(\ref{113}),
 and try to cast it in a form as close as possible to eq.~(\ref{52}):
\be\rlabel{121}
{    \Pi}_S(Q^2) ={ {N_C\Lambda _\chi ^2\over  4\pi ^2G_S}
     {\overline\Pi}_S(Q^2)\over
 {N_C\Lambda _\chi ^2\over  4\pi ^2G_S}-{    \overline\Pi}_S(Q^2) }\ .
\ee
\noindent Using the fact that $G_S^{-1}=\epsilon \Gamma (-1,\epsilon )$,
we can cancel the terms proportional to $\Gamma (-1,\epsilon )$
in the denominator; and write ${    \Pi}_S(Q^2)$ in the simple form
\be\rlabel{122}
{    \Pi}_S(Q^2)=8B_0^2\left\{ 2\tilde{\tilde L_8} +
\tilde{\tilde H_2} +
{2c_m^2(Q^2)\over  Q^2 +M_S^2}\right\}\ ,
\ee
\noindent where
  \be\rlabel{123}
   M_S^2=4M_Q^2\ ,
  \ee
\noindent and ($x_Q={M_Q^2+Q^2x(1-x)\over \Lambda _\chi ^2}$)
\be\rlabel{124}
8B_0^2\times {2c_m^2(Q^2)\over M_S^2} = {N_C\over 16\pi ^2}  4M_Q^2
{{\left( \Gamma (-1,\epsilon )\right)
}^2\over  2\int^1_0 dx \Gamma (0,x_Q)}
= \frac{\langle\bar Q Q\rangle^2}{2M_Q^2 \bar f_\pi^2(Q^2)}\ .
\ee
\noindent The term
 \be\rlabel{125}
8B_0^2(2\tilde{\tilde L_8} + \tilde{\tilde H_2})
 = -{N_C\over 16\pi ^2}4M_Q^2
\Gamma (-1,\epsilon )
 =\frac{\langle\bar QQ\rangle}{M_Q}
\ee
\noindent is not quite the same as in the effective action calculation. The
value of ${    \Pi}_S(Q^2=0)$ however, coincides with the same result
 as in the effective action:
\be\rlabel{126}
{    \Pi}_S(0)=8B_0^2(2L_8+H_2) = {N_C\over 16\pi ^2}  4M_Q^2
{\Gamma (-1,\epsilon )\over 2\Gamma (0,\epsilon )}
\left[ \Gamma (-1,\epsilon )-2\Gamma (0,\epsilon )\right] \ .
\ee
As in the previous subsection,
it was an advantage to reabsorb some of the constant terms into
the pole term,
compared with the expression obtained from the effective action
calculation, to obtain a simple expression to all orders in $Q^2$.

The striking new feature of the summed  scalar propagator is that the
scalar mass is constant and $M_S=2M_Q$. This is to be contrasted with
results of previous work in the literature, see e.g. ref. \cite{21}
and references therein.
Eqs. (\ref{123}) and (\ref{125}) are also true in the presence of gluonic
corrections. They, and the fact that $M_S = 2 M_Q$, are a consequence of
the identities derived in the appendix. It is only the specific
form of the
functions that depends on the inclusion of gluons or not.

\subsection{Two--Point Functions with Mixing }

\quad The case of the other two--point functions ${    \Pi}_A^{(0)}(Q^2)$,
 ${    \Pi}_M^{P}(Q^2)$  and
 ${    \Pi}^{P}(Q^2)$  is somewhat more involved because they mix
 through the Nambu--Jona-Lasinio four--fermion interaction terms
 in eqs.~(\ref{34}) and (\ref{35}).
 Therefore, for this case the result at the n
 bubble level is a matrix equation in terms of the two--point functions
at the n-1 bubble level:
\be\rlabel{127}
{^n\Pi}\equiv\left( \matrix{{^n\Pi}_A^{(0)}\cr {^n\Pi}_M^P \cr {^n\Pi}_P
\cr}\right) = \left( \matrix{
g_V{\overline\Pi}_A^{(0)} & g_S\overline\Pi_M^P & 0 \cr
0 & g_V{\overline\Pi}_A^{(0)} & g_S\overline\Pi_M^P  \cr
0 & g_V{\overline\Pi}_M^{P} & g_S\overline\Pi_P     \cr }\right) \times
\left( \matrix{{^{n-1}\Pi}_A^{(0)}\cr {^{n-1}\Pi}_M^P \cr {^{n-1}\Pi}_P
\cr}\right)\ ,
\ee
\noindent where ${    \overline\Pi}_V^{(0)}(Q^2)$,
${    \overline\Pi}_M^P(Q^2)$
 and
${    \overline\Pi}_P(Q^2)$ denote, as usual the two--point functions at the
one loop level calculated in the mean field approximation; and $g_V$,
$g_S$ are short--hand notation for
\be\rlabel{128a,b}
g_V={8\pi ^2G_V\over N_C\Lambda _\chi ^2} Q^2\,\,\, {\hbox{and}}
\,\,\, g_S= {4\pi ^2G_S\over N_C\Lambda _\chi ^2} \ .
\ee

The series we have to sum, in a vector--like notation, is
\be\rlabel{129}
    \Pi (Q^2)=\sum^{\inf}_{n=1} {^n{\Pi}} =\sum^{\inf}_{n=1}
B^{n-1}(Q^2)\  {\overline\Pi} (Q^2)\ ,
\ee
\noindent where $B(Q^2)$ denotes the $3\times 3$ two--point function
matrix in eq.~(\ref{127}); and the ${    \Pi}(Q^2)$'s three component
two--point function vectors. This series can be summed,
\be\rlabel{130}
{    \Pi}(Q^2)= {1\over  1-B(Q^2)}{^1\Pi}(Q^2)\ .
\ee
\noindent The matrix $(1-B)$ can be inverted, and after some algebra
we get the result
 \be\rlabel{131}
 {    \Pi}_A^{(0)}(Q^2)= {1\over \Delta (Q^2)} \left[
(1-g_S
\overline{    \Pi}^P)\overline{\Pi}_A^{(0)} + g_S{(\overline{\Pi}^P_M)}^2
\right] \ ; \ee
\be\rlabel{132}
{\Pi}^P_M(Q^2)={1\over \Delta (Q^2)}  {\overline\Pi}^P_M(Q^2)\ ;
\ee
\noindent and
 \be\rlabel{133}
 {    \Pi}_P(Q^2)= {1\over \Delta (Q^2)} \left[
(1-g_V\overline{    \Pi}_A^{(0)})\overline{\Pi}_P +
 g_V{(\overline{    \Pi}_M^P)}^2
\right] \ , \ee
\noindent with  $\Delta (Q^2)$  the function
\be\rlabel{134}
\Delta (Q^2)=\left( 1-g_V\overline{    \Pi}_A^{(0)}\right)
 \left( 1 -g_S\overline{    \Pi
}_P\right)   - g_Sg_V{\left( \overline{    \Pi}_M^P\right) }^2\ .
\ee

 It is illustrative to see what the result would be for the scalar
two--point function ${    \Pi}^S(Q^2)$ if we had carried the analysis
keeping the functions $\overline{    \Pi}_V^{(0)}$ and
$\overline{    \Pi}_M^S$:
\be\rlabel{135}
{    \Pi}_S(Q^2)={ -g_V{(\overline{    \Pi}_M^S)}^2 +
(1-g_V\overline{    \Pi}_V^{(0)})
\overline{    \Pi}_S   \over
(1-g_V{    \Pi}_V^{(0)})
(1-g_S\overline{    \Pi}_S)
+g_Sg_V
{(\overline{    \Pi}_M^S)}^2 }\ .
\ee
\noindent If we now set
${\overline    \Pi}_V^{(0)} = {\overline\Pi}_M^S = 0$, we recover the result of
eq.~(\ref{113}).

 We need now to calculate
$\overline{    \Pi}_P(Q^2)$ and
$\overline{    \Pi}_M^P(Q^2)$.
To calculate   $\overline{    \Pi}_P(Q^2)$  we proceed as for the other
 two--point functions we have already calculated. The corresponding
spectral function is
\be\rlabel{136}
{1\over \pi }Im
{    \overline\Pi}_P(t)={N_C\over 16\pi ^2}8M_Q^2{t\over 4M_Q^2}
\sqrt{1-{4M_Q^2\over t}} \theta (t-4M_Q^2)\ .
\ee
\noindent For the large t--behaviour we conclude that
${    \overline\Pi}_P(Q^2)$ obeys a dispersion relation subtracted twice:
\be\rlabel{137}
{    \overline\Pi}_P(Q^2)={    \overline\Pi}_P(0)+Q^2{    \overline\Pi}_P'(0)
+Q^4\int^{\inf}_0{dt\over t^2} {1\over t+Q^2} {1\over \pi }
Im      {\overline\Pi}_P(t) \ ,
\ee
\noindent with the divergent pieces (divergent when
$\epsilon =M_Q^2/\Lambda _\chi ^2\to 0$)
of $     {\overline\Pi}_P(0)$ and
$     {\overline\Pi}_P'(0)$ as known from the effective action calculation
i.e.,
\be\rlabel{138}
{\overline\Pi}_P(0)= {N_C\over 16\pi ^2} 4M_Q^2 \Gamma (-1,\epsilon )\ ;
\ee
\noindent and
\be\rlabel{139}
{    \overline\Pi}_P'(0)=- {N_C\over 16\pi ^2} 2 \Gamma (0,\epsilon )\ .
\ee
The integral in the r.h.s. of eq.~(\ref{137})
has already been calculated in the
section about the scalar  two--point function. We then find the result
($x_Q= {M_Q^2+Q^2x(1-x)\over \Lambda _\chi ^2 }$)
\be\rlabel{140}
{    \overline\Pi}_P(Q^2)= {N_C\over 16\pi ^2}\left\{  4M_Q^2
\Gamma (-1,\epsilon )-2Q^2\int^{1}_0 dx\Gamma (0,x_Q )\right\} \ .
\ee

  Let us check that this result satisfies the three criteria we discussed
in section 3.1. First, there is a Ward identity which relates
 ${\overline    \Pi}_P(Q^2)$  to
 ${\overline    \Pi}_A^{(0)}(Q^2)$, because of the axial current divergence
condition
\be\rlabel{141}
\partial_\mu (\bar Q\gamma ^\mu \gamma _5 Q) = 2M_Q \bar Q i\gamma _5 Q
\ee
in the mean effective field theory. The Ward identity in question, which
we proof in the appendix, is
\be\rlabel{142}
4M_Q^2{\overline    \Pi}_P(Q^2) + 4M_Q <\bar QQ> ={\left( Q^2\right) }^2
{\overline    \Pi}_A^{(0)}(Q^2)\ ;
\ee
\noindent and \cite{12}
\be\rlabel{143}
 <\bar QQ> = -{N_C\over 16\pi^2} 4M_Q^3 \Gamma (-1,\epsilon )\ .
\ee
\noindent Equations (\ref{95}) and (\ref{140}) indeed satisfy this Ward
 identity.
Second, the spectral function calculated from eq.~(\ref{140}) via the
$i\epsilon $ prescription:
\begin{displaymath}
 \log {M_Q^2-tx(1-x)-i\epsilon \over  \Lambda _\chi ^2} =
 \log|{M_Q^2-tx(1-x) \over  \Lambda _\chi ^2}| +i\pi\theta \left(
{tx(1-x)-M_Q^2\over \Lambda _\chi ^2}\right),
\end{displaymath}
is the same as in eq.~(\ref{136}).
 Finally, the first few terms in the $Q^2$--expansion of
 ${    \Pi}_P(Q^2)$ coincide with those calculated in the effective
 action approach with the proper time heat kernel regularization.

The last two--point function to calculate is
 ${    \Pi}^P_M(Q^2)$. The corresponding spectral function is
\be\rlabel{144}
 {1\over \pi } Im     {\overline\Pi}^P_M(t)= {N_C\over 16\pi ^2} 4M_Q
 \sqrt{1-{4M_Q^2\over t}}\theta (t-4M_Q^2) \ .
\ee
\noindent The function  ${    \overline\Pi}^P_M(Q^2)$ obeys a once
subtracted dispersion relation. We already have encountered
the same situation with one of the terms in the scalar two--point
 function, which we have discussed in detail. Therefore, we give the
final result only
\be\rlabel{145}
{    \overline\Pi}^P_M(Q^2)= {N_C\over 16\pi ^2} 4M_Q
 \int^1_0 dx \Gamma (0,x_Q)\ .
\ee

We can now proceed to the explicit calculation of the functions
${\Pi}^{(0)}_A(Q^2)$,
${\Pi}^P_M(Q^2)$ and
${\Pi}_P(Q^2)$ in eqs.~(\ref{131}), (\ref{132}) and (\ref{133}).
We find that
\be\rlabel{146}
\left[ 1- g_S{    \overline\Pi}_P(Q^2)\right]
{    \overline\Pi}_A^{(0)}(Q^2)+ g_S
{\left(     {\overline\Pi}_M^P(Q^2)\right) }^2 = 0 \ ,
\ee
\noindent and hence
\be\rlabel{147}
{    \Pi}_A^{(0)}(Q^2)= 0\ ,
\ee
\noindent a result which must hold in QCD at the chiral limit.
Equation (\ref{146}) also implies that
 \be\rlabel{148}
\Delta (Q^2) = 1- g_S{    \overline\Pi}_P(Q^2) = {Q^2\over M_Q^2}
{1\over 2\Gamma (-1,\epsilon )} \int^1_0 dx \Gamma (0,x_Q) \ ;
 \ee
\noindent and therefore
\be\rlabel{149}
{    \Pi}^P_M(Q^2)=-2 {<\bar Q Q>\over Q^2}  \ .
\ee
\noindent As for ${\Pi}_P(Q^2)$, from eq.~(\ref{133}) and using the
relations
(\ref{146}) and (\ref{148}) above, we find
\be\rlabel{150}
{    \Pi}_P(Q^2)={ {    \overline\Pi}_P(Q^2)-g_V/g_S
{    \overline\Pi}_A^{(0)}(Q^2) \over 1- g_S
{    \overline\Pi}_P(Q^2)         }
=\frac{\langle\bar QQ\rangle}{M_Q}+
\frac{2\langle\bar QQ\rangle^2}{f_\pi^2(Q^2)~Q^2}\ .
\ee
\noindent We see from this result that mixing between the pseudoscalar
 and the longitudinal axial degrees of freedom occurs at all orders in the
 $Q^2$--expansion. Because of eq.~(\ref{148}),
 ${\Pi}_P(Q^2)$ has now a pole
 at $Q^2=0$.
   We want to stress the fact that the final full results in
   eqs.~(\ref{147}),
(\ref{149}) and
(\ref{150}) are very different to those at the one--loop level
approximation. It is only when all the contribution to leading order
in ${1/N_C}$ are summed that these relations, which are expected features
of QCD, appear. These results are a big improvement with respect to the QCD
effective action approach at the mean field approximation.

\subsection{Inclusion of Gluonic Corrections}
\rlabel{gluon}

\quad
In Ref.~\cite{12} the low-energy corrections due to the lowest
dimensional gluonic condensate were also explicitly included.
A general analysis based on the possible types of terms here corresponds
essentially to keeping all the overlined functions as undetermined
parameters but
satisfying the relations derived in the appendix.
This follows from the fact that gluonic lines connecting
different fermion loops in Fig.~2a are suppressed by extra
factors of $1/N_c$ compared to the leading contribution.

The correction due to the leading gluonic vacuum expectation values
can in fact be easily included by using the results for two-point
functions calculated for use in QCD sum rules~\cite{9}.
These corrections can be rewritten in terms of the dimensionless
parameter
\be
\rlabel{ge}
g = \frac{\pi^2}{6N_cm_Q^4}\langle\frac{\alpha_s}{\pi}G^2\rangle
\ee
and the set of functions
\be
\IF_N = \int_0^1 dx\
\frac{1}{\left( 1 + \frac{Q^2}{m_Q^2} x(1-x)\right)^N}
\ .
\ee
The corrections needed for the spin-1 parts are (for $N_c = 3$)
\begin{eqnarray}
\overline{\Pi}_V^{(1)}(Q^2) &=& \frac{3 g M_Q^4}{2\pi^2 Q^4}
\left( -1 + 3\IF_2 -2 \IF_3\right)\ ,
\nonumber\\
\overline{\Pi}_A^{(1)}(Q^2) &=& \frac{9 g M_Q^4}{2\pi^2 Q^4}
\left( 1-\IF_2\right)\ .
\end{eqnarray}
For the scalar two-point function we need
\be
\overline{\Pi}_S(Q^2)  =  \frac{9 g M_Q^4}{4\pi^2 Q^4}
\left( -1-2\IF_1 + 3\IF_2\right)\ .
\ee
The spin-0 axial-vector, pseudo-scalar and mixed two-point functions have
very large cancellations in the denominator and are numerically very
unstable when gluonic corrections are included. They can be handled
similarly in principle. The gluonic correction terms do also satisfy the
relations derived in the appendix as required.

We have checked that using the above formulas the two-point functions
including non-zero gluonic vacuum expectation values converge for
small values of $Q^2$ to the low energy expansion with these
corrections included; i.e. those of eqs. (21)-(28) with the values of the
parameters calculated including gluonic corrections.
All the nice features of the two-point functions as given in
the previous subsections are still valid since the underlying cause
for these properties were the relations derived in the appendix and the
gluonic corrections have to satisfy those as well.

\subsection{Numerical Results}
\rlabel{numeric}

\quad In this section we plot the two-point functions as calculated in the
previous subsections.
As described in subsection \ref{gluon} we can also include
gluonic corrections. In view of the result of ref. \cite{12} that a very
good fit to the low-energy parameters could also be obtained
without gluonic corrections, we only show the
effect of the gluonic corrections in the vector two-point functions.
The input values used for all of the plots are those of
fit 1 in ref. \cite{12}.
They are $M_Q = 0.265~GeV$, $\Lambda_\chi = 1.165~GeV$ and $g_A = 0.61$.
The variation with the input parameters can be judged from table 2 in
ref. \cite{12}. The size of the changes here is similar to the ones obtained
there at values of momentum transfer $Q^2 = 0$.

In fig. 4 we have plotted the vector-two-point function for positive
values of $Q^2$ for $\sqrt{Q^2}$ from 0 to 1.5 $GeV$. The full line
corresponds to $\overline{\Pi}_V^{(1)}(q^2)$ in the full ENJL--model.
The dashed line is
the corresponding result using the effective approximation, eq. (\ref{54}).
The vector meson mass for the values of the
parameters $M_Q$, $\Lambda_\chi$ and $g_A$ which we have fixed,
 is about 0.81~$GeV$.
The short-dashed line corresponds to the
vector-two-point function in the QCD
effective action model of ref. \cite{14}. As can be seen
in the figure, the
full resummation leads to lower values for the two point
functions than those of the
low-energy formulas when extended to higher $Q^2$. In the same figure 4 we
also show the effect of the gluonic corrections. The dotted line
uses the same input values as given above but has a non-zero value for
the gluonic background,
we have set $g = 0.5$ (see eq. (\ref{ge})). It can be seen that the effect
of gluonic corrections is large at
small values of $Q^2$ but grows smaller
at higher values of $Q^2$.

In fig. 5 we have plotted similarly to the vector case, the result for
$Q^2 \Pi^A_1(Q^2)$.
(The extra factor of $Q^2$ is included to remove the pole
at $Q^2 = 0$ due to pion exchange.)
In contrast to the vector two-point function,
there is also a significant difference between the one-loop result of
eq. (\ref{55})
(dashed line) and the full resummation of eq. (105) at
low $Q^2$(the full line).
 This is
due to what is usually called the pseudoscalar-axial-vector mixing
and was in our previous work described by the coupling $g_A$. The value of
this two-point function
at $Q^2 = 0$ determines $f_\pi^2$. The dashed line
corresponds to the
two-point function using the effective approximation of eq. (\ref{55}).
The axial-vector mass here is about 1.3~$GeV$.

Fig. 6 shows how the summation of the whole series of
diagrams has produced the pole at
$Q^2 = 0$ that is required by the spontaneous
breaking of the chiral symmetry
for the pseudoscalar two-point function.
The one-loop result (dashed line)
 does not
have this behaviour, but the full resummed version (full line) of
eq. (\ref{150})
does. The short-dashed line is the low-energy extrapolation of eq. (28).
Here we see how the full resummation correctly reproduces the low-energy
behaviour as derived in ref. \cite{12}, but for larger values of momentum
transfer, it starts to differ appreciably.

We have not plotted the scalar two-point function. The pole is, as we have
proven above, always at $M_S = 2 M_Q$.
This pole is generated by the full
resummation.
Neither have we plotted the mixed pseudoscalar--axial-vector two-point
function since this has the very simple behaviour of Eq. (\ref{149}).

The overall picture of the
high energy behaviour of the ENJL--model that emerges
after the resummation, is
improved
compared to the behaviour obtained from the low-energy expansion.
This is illustrated
by the fact that now it also satisfies the second Weinberg
sum rule.
The ENJL--model has another advantage over the simple
QCD effective action model of ref. \cite{14}.
By virtue of the extra 4-quark interactions present, this model naturally
contains
more or less correct meson poles while the simple quark version, that
corresponds
to the one-loop result (or essentially the use of the overlined two-point
functions), does not.
This means that for positive values of $q^2$ the two-point
functions are considerably enhanced both in the real and imaginary parts
as compared to the one loop result.
The importance of this type of behaviour
can be seen, e.g. in the determination of some low-energy constants
using dispersion relations.  As an example we show in fig. 3 how the
imaginary part of the vector two-point function gets enhanced
considerably over the one-loop result.

The advantage of
using the full ENJL--model over a parametrization with meson resonances
is that the number of free parameters
remains within limits. For instance,
if one tries to extend the analysis of ref. \cite{15} to non-leptonic
matrix elements
using a parametrization with vector mesons it requires the
knowledge of weak decays of vector mesons which have not been observed
 experimentally.

In ref. \cite{12} the masses and couplings of the mesons were determined
from the low-energy expansion. These are essentially given by
various
combinations of derivatives of the two-point functions at $q^2 =0$.
An alternative way of determining
the meson masses is to determine them by
looking for the poles and residues of the full two-point functions.
This procedure can be questioned on
the grounds that for euclidean momenta
quark confinement is not so important
but for momenta of $q^2 \ge 4 M_Q^2$ we
get effects of free quarks included. The two-point functions still have
poles though. As an example we give the position of the poles for the
vector, axial-vector and scalar for the parameters used above, in
table \ref{table1}. The pion mass is of course exactly zero in both cases
since we work in the chiral limit.
In general the masses are lower than those derived from the
low energy approximation.

\section{ THE $\pi^+$--$\pi^0$ ELECTROMAGNETIC MASS DIFFERENCE}
\quad To the lowest order in the chiral
 expansion, the effect of virtual
electromagnetic interactions to lowest order in the fine structure
constant $\alpha _{em}=e^2/4\pi $, generates a term in the effective
action without derivatives \cite{22}:
\be\rlabel{151}
\int d^4 x\left\{ e^2 C_1 tr QU(x)QU^+(x)\right\} \ ,
\ee
\noindent where $U$ is the unitary matrix which collects the
pseudoscalar Goldstone fields and $Q$ the quark electric charge matrix
$Q=1/3 \mathrm{diag}(2,-1,-1)$. Expanding eq.~(\ref{151})
 in powers of pseudoscalar
fields
\be\rlabel{152}
 e^2C_1 tr QUQU^+ = -{2e^2C_1\over  f_\pi ^2 }(\pi^+\pi^- +K^+K^-)
+ {\cal O}(\phi^4)\ ,
\ee
\noindent one sees explicitly that this term leads to a
$\pi^+$--$\pi^0$ (and $K^+$--$K^0$) mass splitting
\be\rlabel{153}
\Delta m_\pi^2 ={\left( m_{\pi^+}^2-m_{\pi^0}^2\right)}_{EM}
 = {2e^2C_1\over f_\pi ^2}\ .
\ee
 \noindent The constant $C_1$, like $f_\pi ^2$, is not fixed by symmetry
requirement alone. It is determined by the dynamics of the underlying
theory. Formally, it is given by the integral representation \cite{13}
\be\rlabel{154a}
 2e^2C_1=-ie^2\int{d^4q\over (2\pi )^4}
 {g_{\mu \nu } -{q_\mu q_\nu \over q^2} \over q^2 -i\epsilon }
(q^\mu q^\nu - q^2g^{\mu \nu } ){\Pi}_{LR}^{(1)}(q^2)\ ,
\ee
\noindent where
\be\rlabel{154b}
{\Pi}_{LR}^{(1)} = {1\over 2}\left( {\Pi}_V^{(1)} -{\Pi}_A
 ^{(1)}\right) \ ,
\ee
\noindent with
${\Pi}_V^{(1)}$ and ${\Pi}_A^{(1)}$ the vector and axial vector
invariant functions we have discussed. Performing a Wick rotation
in eq.~(\ref{154a}) leads to the sum rule \cite{2}
\be\rlabel{155}
\Delta m_\pi ^2 ={\alpha _{em}\over \pi} {1\over 16\pi ^2f_\pi ^2}
(-6\pi ^2)\int^\inf_0 {dQ^2\over Q^2} Q^4\left[
{\Pi}_V^{(1)}(Q^2)-{\Pi}_A^{(1)}(Q^2)\right] \ .
\ee

The purpose of this section is to discuss the evaluation of the
$\Delta m_\pi ^2$ sum rule above based on the two--point function
results discussed in the previous sections. It is then convenient
 to split the $Q^2$--integral into long--distance
($0\le Q^2 \le \mu^2$) and short--distance ($\mu^2\le Q^2\le\infty$)
parts:
\be\rlabel{156}
\int^{\inf}_0dQ^2\cdot \cdot \cdot =
\int^{\mu^2}_0dQ^2\cdot \cdot \cdot +
\int_{\mu^2}^{\inf} Q^2\cdot \cdot \cdot
\ee
\noindent We shall concentrate first on the long--distance part
calculation.

{\bf a) Long--distance contribution. Phenomenological
   approach}

The very low $Q^2$ contribution to the integral
\be\rlabel{157}
{\left( \Delta m_\pi ^2\right) }_{LD} = {\alpha _{em}\over \pi}
\left( -3\over 8f_\pi ^2\right)  \int^{\mu^2}_0 dQ^2  Q^2\left(
 {\Pi}_V^{(1)}-{\Pi}_A^{(1)}\right)
\ee
is fixed by chiral perturbation theory (see eqs.~(\ref{21}) and
(\ref{23})):
\be\rlabel{158}
{\Pi}_V^{(1)}(Q^2)-{\Pi}_A^{(1)}(Q^2)= {-2f_\pi ^2\over Q^2}
-8L_{10} + {\cal O}(Q^2)\ ,
\ee
\noindent from which it follows that \cite{13}
\be\rlabel{159}
\left( \Delta m_\pi ^2\right)_{\chi PT}= {\alpha _{em}\over \pi }
{3\over 4} \mu^2 \left\{ 1 + {2L_{10}\over f_\pi ^2}\mu^2 +{\cal O}(
\mu^4)\right\} \ .
\ee
\noindent The known correction term ${\cal O}(\mu^2)$ in the parenthesis
of the r.h.s. can be used to estimate the value of the $\mu^2$--scale
 at which we can trust the validity of the $\chi$PT--contribution.
 {}From the fact that \cite{20}
\be\rlabel{160}
{4L_{10}\over f_\pi ^2} \simeq -{1\over M_\rho^2}\ ,
\ee
\noindent we conclude that the $\chi $PT--result in
eq.~(\ref{159}) can only
 represent correctly the long--distance contribution to $\Delta m_\pi ^2$
up to scales
\be\rlabel{161}
 \mu^2_{\chi PT}< M_\rho^2  \ .
\ee
\noindent Obviously, this is too small a scale to trust numerically
 a direct matching with the short--distance contribution, which as we
shall see later, it is expected to be valid for $\mu^2$--scales larger
than a few $GeV^2$ at least. (See however
the first paper in ref.\cite{22} and ref. \cite{13}.)

      Since the early work of Das et. al. \cite{2}, the traditional
phenomenological approach to the calculation of $\left( \Delta m_\pi ^2
\right)_{LD}$ has been to include the effect of vector and axial--vector
particle states in the $Q^2$--integral, using a parametrization that is
constrained to satisfy the first and second Weinberg sum rules.
The usual phenomenological VMD--model parametrization is
\be\rlabel{162}
\Pi_V^{(1)}={2f_V^2M_V^2\over M_V^2+Q^2}
\ee
\noindent and
\be\rlabel{163}
\Pi_A^{(1)}={2f_\pi ^2\over Q^2}+{2f_A^2M_A^2\over M_A^2+Q^2}\ ,
\ee
\noindent with the constants $f_\pi ^2$, $f_V^2$, $f_A^2$, $M_V^2$
and $M_A^2$ constrained
by the relations
\be\rlabel{164a}
f_\pi^2 + f_A^2M_A^2 = f_V^2M_V^2
\ee
\noindent and
\be\rlabel{164b}
f_V^2M_V^4 = f_A^2M_V^4,
\ee
\noindent which ensure the convergence of the limits
\be\rlabel{165a,b}
\lim_{Q^2\to \inf} Q^2(\Pi_V^{(1)}-{\Pi}_A^{(1)})\to 0
\,\,\, {\hbox{and}} \,\,\,
\lim_{Q^2\to \inf} Q^4(\Pi_V^{(1)}-{\Pi}_A^{(1)})\to 0\ ;
\ee
\noindent i.e., the superconvergence relations which lead to the first
 and second Weinberg sum rules. One then has
 \be\rlabel{166}
 \left( \Delta m_\pi ^2\right)_{VMD}={\alpha _{em}\over \pi }
 {3\over 4} \int^{\mu^2}_0 dQ^2 {M_A^2M_V^2\over  (Q^2+M_A^2)(Q^2+M_V^2)}
 \ee
 \noindent For $M_A$, $M_V$ $\rightarrow  \inf$, with $\mu^2$ fixed
 we recover the first term of the $\chi$PT calculation in
 eq.~(\ref{159}).
 If we let the
 scale $\mu^2$ go to infinity; then, for $M_A=\sqrt{2}M_V$, one finds
 the early result of Das et al. :
 \be\rlabel{167}
 \left( \Delta m_\pi ^2\right) _{\cite{2}}={\alpha _{em}\over \pi }
 {3\over 2}M_\rho ^2log2=1.4\times {10}^3MeV^2\ .
\ee
\noindent Experimentally,
\be\rlabel{168}
\left( m_{\pi ^+}-m_{\pi ^0}\right)_{Exp.} = (4.5936\pm 0.0005)MeV\ ,
\ee
\noindent while the phenomenological result of Das et al. corresponds to
\be\rlabel{169}
\left( m_{\pi ^+}-m_{\pi ^0}\right)_{\cite{2}} = 5.2MeV\ ,
\ee
\noindent Recent  phenomenological evaluations of the $\Delta m_\pi ^2$
sum rule, which include explicit chiral symmetry  breaking effects,
can be found in refs.\cite{23} to \cite{25}.

{\bf b) Long--distance contribution in the ENJL--model}

  The calculation of $\left( \Delta m_\pi ^2\right) _{LD}$  in the
QCD effective action approach of ref. \cite{14},
which corresponds to the mean
field approximation of the Nambu Jona-Lasinio model, was reported in
ref.\cite{13}.   It is the approximation where
 \be\rlabel{170}
{\Pi}_V^{(1)}-{\Pi}_A^{(1)}\to
{\overline\Pi}_V^{(1)}-{\overline\Pi}_A^{(1)} = -{N_C\over 16\pi ^2}
{8M_Q^2\over Q^2} \int^1_0 dx \Gamma (0,x_Q )\ ,
\ee
\noindent which is the result obtained in eqs.~(\ref{90}) and (\ref{95}).
This leads
to the result $(\epsilon =M_Q^2/     \Lambda _\chi ^2,
x_Q={Q^2x(1-x)+M_Q^2\over \Lambda _\chi ^2})$
\be\rlabel{171}
 \left( \Delta m_\pi ^2\right)_{\cite{13}}={\alpha _{em}\over \pi }
 {3\over 4} \int^{\mu^2}_0 dQ^2 {1\over \Gamma (0,\epsilon )}
 \int^{1}_{0}dx\Gamma (0,x_Q)\ .
 \ee
 \noindent In ref.\cite{13} a proper time regularization for the photon
 propagator was used; and for simplicity, the $\mu^2$--scale was
 identified with $\Lambda _\chi ^2$.
 The shape of this mean field approximation evaluation versus $\mu^2$,
 for the input value of $M_Q^2$ and $\Lambda _\chi ^2$  which we have
 been considering (i.e., the value corresponding to fit 1 in ref.\cite{12}:
 $M_Q = 265 MeV$ and $\Lambda _\chi  = 1165 MeV$) is plotted in Fig.~7.

   The evaluation of $(\Delta m_\pi ^2)_{LD}$
 in the full ENJL--model, with the expressions of the two--point
 functions ${\Pi}_V^{(1)}(Q^2)$ and   ${\Pi}_A^{(1)}(Q^2)$
obtained in the previous section leads to the result
 \be\rlabel{172}
 \left( \Delta m_\pi ^2\right)_{ENJL}={\alpha _{em}\over \pi }
 {3\over 4} \int^{\mu^2}_0 dQ^2 {f_\pi ^2(Q^2)\over f_\pi ^2}
 {M_A^2(Q^2)M_V^2(Q^2)\over  \left( Q^2+M_A^2(Q^2)\right)
 \left( Q^2+M_V^2(Q^2)\right) } \ ,
 \ee
\noindent with the $Q^2$--dependent functions
 $M_V^2(Q^2)$, $M_A^2(Q^2)$  and  $f_\pi ^2(Q^2)$  as given by
 eqs.~(\ref{81}),
(\ref{105}), (\ref{102}) and (\ref{104}).

 The shape of $(\Delta m_\pi ^2)_{ENJL}$  versus $\mu ^2$  is shown in
 Fig.~7. We expect the integrand in eq.~(\ref{172}) to be a good
 representation of the low and intermediate energy scales; and
 therefore, the matching with short--distance evaluation should now be
much smoother than in the case of the mean field approximation. This we
discuss in the next subsection.

{\bf c) Short--distance contribution and numerical
results.}

\quad In QCD perturbation theory
${\Pi}_V^{(1)}(Q^2) = {\Pi}_A^{(1)}(Q^2)\ .$
Spontaneous symmetry breaking induces a deviation from this result which,
at large $Q^2$ and to leading order in the $1/N_C$--expansion, can be
calculated using the operator product expansion, with the result
(\cite{8} and first ref. in \cite{22})
\be\rlabel{173}
 \left( {\Pi}_V^{(1)}-{\Pi}_A^{(1)}\right) = -{1\over Q^6}
{3\pi ^2\over 2} {N_C\alpha _s(Q^2)\over \pi }{(<\bar\psi\psi>)}^2\ ,
\ee
\noindent where $(N_C\to \inf )$:
\be\rlabel{174}
{N_C\alpha _s (Q^2)\over \pi }\to {6\over
11 \log({Q\over \Lambda _{QCD}})}
\ ;
\ee
\noindent and
\be\rlabel{175}
<\bar\psi\psi(Q^2)>={\hat{<\bar\psi\psi>}} {\left(
\log (Q/\Lambda _{QCD}) \right)}^{9\over 22} \ .
\ee
\noindent Inserting this asymptotic estimate in the short--distance
expression for $\Delta m_\pi ^2$, leads to the result
 \be\rlabel{176}
 \left( \Delta m_\pi ^2\right)_{SD}={\alpha _{em}\over \pi }
 {27\pi ^2\over 88f_\pi ^2} \left({\hat{<\bar\psi\psi>}}
 \over \mu^2\right)^2
\int^{\inf}_1 {dz\over z^2} \left( {1\over 2}\log \left( {\mu ^2\over
\Lambda _{QCD}^2}z \right)\right)^{-{2\over 11}}\ .
 \ee

Fig.7 also shows the shape of $(\Delta m_\pi ^2)_{SD} $ versus
$\mu^2$ for various values of the invariant quark condensate
$\hat{<\bar\psi\psi>}$. \footnote{The continuous curve is the one
corresponding to the choice $|<\bar\psi\psi>| = (281 MeV)^3$, which
is the value predicted in the ENJL--model for the input values $M_Q=265
MeV$ and $\Lambda _\chi =1162 MeV$. }
Obviously, as the scale $\mu^2$ becomes small
$\left( \Delta m_\pi ^2\right) _{SD}$ diverges. The matching between
$\left( \Delta m_\pi ^2\right) _{SD}$ and
$\left( \Delta m_\pi ^2\right) _{LD}$ is defined by the optimal choice
of $\mu^2$ which minimizes the variation of the total $\Delta m_\pi^2$.
As seen in fig. 8 this occurs at value
$\mu \approx 950~MeV$;
and in fact around the value, the stability is rather good.
The corresponding value of $\Delta m_\pi^2$ in this range, is
\be\rlabel{????}
         \Delta m_\pi^2 \approx 1.3\cdot 10^{-3} ~GeV^2\ ,
\ee
\noindent and agree well with the experimental value, the horizontal
dashed line in fig. 8.
\section{CONCLUSIONS}

\quad
In this paper we have extended the general analysis of the ENJL--model
as done in ref. \cite{12} beyond the low-energy expansion.
We have calculated directly the two-point functions within the ENJL--model
to all orders in momenta. The relations that the one-loop results have to
satisfy lead after the full resummation to a set of rather simple forms
for the two-point functions.
It should be stressed once more that these are satisfied independent of
the gluonic
interactions and are thus valid in a wide class of ENJL-like models.

The resulting expressions are,
 for the vector-axial-vector cases, very similar
 to the ones
usually obtained
 assuming some kind of vector, axial-vector meson dominance.
The full resummations have a well
behaved high-energy behaviour. They satisfy
both the first and the second Weinberg sum rules.
The resummation also
obeys the Ward identities of the full theory.

Simple expressions were also found
for the other two-point functions. A byproduct
was a proof that within this
class of models the scalar two-point function
always has a pole corresponding
to a mass of twice the constituent quark mass.
Our derivation only depends on the underlying symmetry properties of the
Lagrangian and is hence regularization scheme independent.
The full resummation also reproduced the pole at $Q^2=0$ in the
pseudo-scalar two-point function explicitly showing how this
model obeys the Goldstone theorem.

Finally, the two-point functions derived were
used to start evaluating nonleptonic matrix
elements within the class of
ENJL-like models. We have estimated the electromagnetic
$\pi^+-\pi^0$ mass difference and found good agreement with the measured
value.

\section*{ACKNOWLEDGEMENTS}
We would like to thank Ch. Bruno for collaboration in the early stages
of this work and helpful comments. H.Z. would like to thank the 
ICSC world laboratory for financial support. J.B. thanks CPT Marseille
for hospitality. 
%
% appendix with the derivation of the Ward-Identity.
\appendix
\section*{APPENDIX}

\quad In this appendix we derive the Ward
identities that the one-loop two-point
functions have to satisfy. We first
give a derivation based on the heat-kernel
expansion and a general analysis of
the type of terms that can contribute to
the two-point functions. This method allows for explicit contact to be
made with the regularization chosen in the heat-kernel expansion.
A second method is essentially the
traditional way of deriving Ward identities
but we have to take into account that
$\langle \overline{q}q\rangle \ne 0$.
The second
method can also be used to derive some of the identities that the
full two-point functions have to satisfy.

The one-loop two-point functions are calculated using the Lagrangian
($U=1$)
\begin{equation}
\rlabel{eqX1}
{\cal L} = \overline{q} i {D\hspace{-1.2ex}/\ } q - M_Q \overline{q}q
-\overline{q}(s-ip\gamma_5)q = \overline{q}{\cal D}q \ .
\end{equation}
The last equality is the definition of ${\cal D}$ and the covariant
derivative
$D\hspace{-1.2ex}/\ $ contains the vector and axial-vector external
fields. The real part of the
effective action in Euclidean space using the heat kernel expansion
is then given by ($\epsilon = M_Q^2 / \Lambda_\chi^2$):
\begin{equation}
S_{eff} = -\frac{1}{32\pi^2}
\sum_{n\ge 1} \Gamma(n-2,\epsilon) (M_Q^2)^{2-n}\int d^4 x \tr
{\cal H}_n(x)\ .
\end{equation}
The ${\cal H}_n(x)$
are the Seeley-DeWitt coefficients and these are constructed
out of $E,\ R_{\mu\nu}$ and their covariant derivatives. These are defined
by
\begin{equation}
{\cal D}_E^{\dagger} {\cal D}_E = - D_\mu D^\mu + E + M_Q^2
\qquad \mathrm{and}\qquad \left[ D_\mu , D_\nu \right] = R_{\mu\nu} \ .
\end{equation}
In terms of the external fields $s,\ p,\ l_\mu$ and $r_\nu$ they are
(only terms that can contribute to two-point functions are given):
\begin{eqnarray}
E & = &  i\gamma_\mu\gamma_5 M_Q \left(r_\mu - l_\mu\right)
-\frac{i}{2}\sigma_{\mu\nu}R_{\mu\nu}\nonumber\\&&
+ s^2 + M_Q s + p^2 +\gamma_\mu\partial_\mu s
-i\gamma_\mu\gamma_5\partial_\mu p\ ,                  \\
\rlabel{eqX6}
R_{\mu\nu} & = & -\frac{i}{2}\left(
l_{\mu\nu}+r_{\mu\nu}-\gamma_5\left(l_{\mu\nu}-r_{\mu\nu}\right)\right)
             \ .
\end{eqnarray}
Here we see that $E$ and $R_{\mu\nu}$ vanish for vanishing external
fields so
only terms containing at most two factors of $E$ and $R_{\mu\nu}$
can contribute to the two-point functions.

The first two coefficients are:
\begin{equation}
{\cal H}_0 = 1 \qquad\mathrm{and}\qquad {\cal H}_1 = -E\ .
\end{equation}
${\cal H}_1$
thus contributes to the scalar and pseudoscalar two-point function.
These are
the only two-point functions that contain a quadratic divergence.

The ${\cal H}_{n\ge 2}$
only contain two types of terms that can contribute to
two-point functions. Let us look at all possibilities.

Terms with a single $E$. These are of the form $D^{2(n-1)}E$ and are
total derivatives, so they
do not contribute to the two-point functions. The same
argument applies to terms with a single $R_{\mu\nu}$.

Terms with one $E$ and one $R_{\mu\nu}$. Extra derivatives acting
on these can always be commuted, the commutator introduces an extra
factor of
$R_{\mu\nu}$ and then only contributes earliest to a three point
function. We can also use partial integration. All this type of terms can
thus be brought into the form
\begin{equation}
D_\mu D_\nu E D^{2(n-3)} R_{\mu\nu} =
\frac{1}{2}\left[D_\mu , D_\nu\right] E D^{2(n-3)}R_{\mu\nu} \ .
\end{equation}
The commutator
becomes an extra factor of $R_{\mu\nu}$ so this type of terms
does not contribute to two-point functions. We conclude that
the ${\cal H}_{n\ge 2}$
only contribute to two-point functions through terms
like
\begin{equation}
E D^{2(n-2)}E\ ,
\qquad D_\alpha R_{\alpha\beta}D^{2(n-3)}D_\mu R_{\mu\beta}
\qquad {\rm and}\qquad R_{\alpha\beta}D^{2(n-2)}R_{\alpha\beta}\ .
\end{equation}
Using Eq. (\ref{eqX6}) the last two terms are of the form
\begin{equation}
D_\alpha v_{\alpha\beta}D^{2(n-3)}D_\mu v_{\mu\beta} +
D_\alpha a_{\alpha\beta}D^{2(n-3)}D_\mu a_{\mu\beta}
\end{equation}
and
\begin{equation}
v_{\alpha\beta}D^{2(n-2)}v_{\alpha\beta}+
a_{\alpha\beta}D^{2(n-2)}a_{\alpha\beta} \ .
\end{equation}
So these contribute only to the transverse part and equally for the
vector and the axial-vector two-point function. The first term has a part
containing $R_{\mu\nu}$ as well. It contributes only to the
transverse part, and equally for the vector and axial-vector case.

The remaining type of
terms can be rewritten using the explicit form of $E$.
\begin{eqnarray}
\int d^4x \tr E D^{2(n-2)}E &=& N_c\int d^4x \tr \big[
16 M_Q^2 A_\mu\partial^{2(n-2)}A_\mu
+ 16 M_Q A_\mu \partial^{2(n-2)}\partial_\mu P \nonumber\\
+ 16 M_Q^2 P \partial^{2(n-1)}P
&+& 16 M_Q^2 S\left( 1 +\frac{ \partial^2}
{4M_Q^2} \right) \partial^{2(n-2)} S
\big]\ .
\end{eqnarray}
The axial-vector terms contribute only proportionally to $g_{\mu\nu}$. This
together with the above contribution leads to:
\begin{eqnarray}
\rlabel{eqX12}
\overline{\Pi}^{(0)}_V(Q^2) &=& \overline{\Pi}_{(M)}^S(Q^2) = 0 \ ,\\
\overline{\Pi}^{(1)}_V(Q^2) &=&
\overline{\Pi}^{(1)}_A(Q^2) + \overline{\Pi}^{(0)}_A(Q^2)\ .
\end{eqnarray}
The first two of these equations appear because the vector current in the
Lagrangian Eq. (\ref{eqX1}) is conserved. The third one is the reason why
the first Weinberg sum rule is satisfied even at the one-loop level.
It also guarantees both Weinberg sum rules after the resummation.
Including the contributions from ${\cal H}_1$ we also have
\begin{eqnarray}
\rlabel{eqX14}
-2 M_Q \overline{\Pi}_{M}^P(Q^2) &=&
Q^2 \overline{\Pi}^{(0)}_A(Q^2) \ ,\\
2 M_Q \overline{\Pi}_P(Q^2) &=& -2\langle\overline{Q}Q\rangle -
Q^2 \overline{\Pi}^P_M(Q^2)
\rlabel{eqX16} \ ,\\
\overline{\Pi}_S(Q^2) &=&
\overline{\Pi}_P(Q^2) + Q^2\overline{\Pi}^{(0)}_A(Q^2)
\rlabel{eqX17} \ .
\end{eqnarray}
In Eq. (\ref{eqX16}) we have used the relation between the coefficient
of ${\cal H}_1$
and the quark vacuum expectation value. In the chiral limit this
vacuum
expectation value is determined uniquely by the contribution of
${\cal H}_1 = -E$.
This derivation is also valid in the presence of low-frequency gluons.
The effective
action after including the low-energy gluonic effects through
gluonic vacuum
expectation values, still has to be constructed out of $E$ and
$R_{\mu\nu}$. This was precisely the argument used in Ref. \cite{12} to
obtain relations
between the low-energy coupling constants that are independent
of the gluonic corrections. The results following from the relations
(\ref{eqX12}-\ref{eqX17})
after resummation are the equivalent relations for
the two-point functions. This is what we used in Sect. 3 to rewrite
all the two-point functions in terms of essentially
two functions and one constant.

The preceding derivation was obtained using the Seeley-DeWitt
expansion to all orders. Let us  now show how several results
can also be obtained from the underlying relations
in the Lagrangian (\ref{eqX1}). These relations are:
\begin{eqnarray}
\partial_\mu\left(\overline{q}\gamma_\mu q\right) &=& 0 \ ,\\
\partial_\mu\left(\overline{q}\gamma_\mu\gamma_5 q\right)
&=& 2 i M_Q \overline{q}\gamma_5q
 \ ,\\
\{q^{a\dagger}_\alpha(x) , q^b_\beta (0)\} &=&
\delta^{ab}\delta_{\alpha\beta} \delta^3(x)\ .
\rlabel{eqX19}
\end{eqnarray}
Eq. (\ref{eqX19})
is valid at equal times. $a,b$ are colour-flavour indices and
$\alpha,\beta$ are Dirac spinor indices.

We start from
\begin{eqnarray}
q_\mu \overline{\Pi^A_{\mu\nu}} &=&
\int d^4x \left(\partial_\mu e^{iq\cdot x}\right)
\langle 0 | T \left( A_\mu(x) A_\nu(0) \right) | 0 \rangle \\
&=& -2M_Q
\int d^4x \langle 0 | T \left( P_\mu(x) A_\nu(0) \right) | 0 \rangle
\nonumber\\&&
-\int d^4x \delta_{\mu 0}\delta(x^0)
\langle 0 |  \left[ A_\mu(x), A_\nu(0) \right] | 0 \rangle \\
&=& -2iM_Q\overline{\Pi^{(P)}_\nu}\ .
\end{eqnarray}
The matrix element of the equal time commutator
vanishes for two identical currents. This follows
from Eq. (\ref{eqX19}). Putting in the form of the two-point functions
this leads to
\begin{equation}
q^2 q_\nu \overline{\Pi^{(0)}_A} = 2 M_Q q_\nu \overline{\Pi^P_{(M)}}
\end{equation}
or the same as equation (\ref{eqX14}). In the full theory we have
$\partial_\mu A_\mu = 0$ so the identical derivation leads to:
\begin{equation}
\Pi^{(0)}_A (Q^2) = 0\ .
\end{equation}
This  equation is satisfied by the fully resummed two-point function.

A similar derivation leads to
\begin{equation}
q_\mu \overline{\Pi^P_\mu} = 2i M_Q \overline{\Pi^P}
 -\int d^4x \delta_{\mu 0}\delta(x^0)
\langle 0 |  \left[ A_\mu(x), P(0) \right] | 0 \rangle\ .
\end{equation}
Here the equal
time commutator worked out using Eq. (\ref{eqX19}) does not
vanish. A term
proportional to the quark vacuum expectation value remains and
leads to
Eq. (\ref{eqX16}). In the full theory $\partial_\mu A_\mu = 0$ so
we obtain
\begin{equation}
\Pi^P_{(M)}(Q^2) = -2\frac{\langle\overline{Q}Q\rangle}{Q^2}\ .
\end{equation}
This equation is also satisfied by the fully resummed two-point function.

\newpage
\begin{table}
\caption{Values of the masses determined from the poles in the two-point
functions and from the low-energy expansion of ref. \protect{\cite{12}}.}
\rlabel{table1}
\begin{center}
\begin{tabular}{|c|c|c|}
\hline
Meson & ref. \protect{\cite{12}} & Pole \\
\hline
$M_V$   & 0.81~$GeV$    &  0.70~$GeV$       \\
$M_A$   & 1.3~$GeV$     &  0.9~$GeV^{\dagger}$       \\
$M_S$   & 0.62~$GeV$    &  0.53~$GeV$ $= 2M_Q$\\
\hline
\end{tabular}
\end{center}
${}^{\dagger}$ In the resummed version there is an strong
enhancement around this
value of the two-point function. It does not become a pole with the values
of the parameters chosen here.
\end{table}

\newpage
\section*{Figure Captions}
\newcommand{\doos}[1]{\parbox[t]{0.9\textwidth}{\noindent #1}}
\newcommand{\doosa}[1]{\parbox[t]{0.80\textwidth}{\noindent #1}}

Fig. 1: \doos{(a): \doosa{The set of diagrams summed to obtain $g_A(Q^2)$.
X is the insertion of the pion field and the other lines are fermions.}\\
(b): \doosa{The gap equation. The thick line is the full fermion propagator.
The thin line is the bare fermion propagator.}}\\
Fig. 2: \doos{(a): \doosa{The set of diagrams to be summed for the two-point
functions.}\\
(b): \doosa{The one loop fermion bubble.}}\\
Fig. 3: \doos{The spectral function
of the vector two-point function. The full
line is the full ENJL result. The dashed line is the result at one-loop. Here
$Q=\sqrt{t}$.}\\
Fig. 4: \doos{The real part of the vector two-point function. Plotted are
the full result with
(labelled gluon) and without
(labelled full) gluonic corrections. The VMD parametrization (eff)
and the one loop result (1-loop).}\\
Fig. 5: \doos{The real part of the axial-vector two-point function multiplied
 by $Q^2$. The effect of axial-pseudoscalar mixing that the full resummation
reproduces is visible at all $Q$'s. The labels have the same meaning
as in Fig. 4.}\\
Fig. 6: \doos{The real part of the pseudoscalar two-point function
multiplied by $Q^2$. Notice how
the resummed version produces the pole at $Q=0$. The labels have the same
meaning as in Fig. 4.}\\
Fig. 7: \doos{Curves for $\Delta m_\pi^2$ in terms of the scale $\mu$.
Plotted are the long-distance part for the pion exchange term only (LD-CHPT),
the result of ref. \cite{13}
(LD-mean) and the ENJL-result after the resummation
(LD-ENJL). The short distance contributions are plotted for three values
of $\langle\bar Q Q\rangle = -(194~MeV)^3$(SD194), $-(220~MeV)^3$ and
$-(281~MeV)^3$.       }\\
Fig. 8: \doos{The full result $\Delta m_\pi^2$ versus $\mu$ corresponding
to the sum of the long distance ENJL result with the short distance
evaluation using
$\langle\bar Q Q\rangle $ as given by the ENJL--model  (see text).}

\newpage

%
% F:\TEX\TEXDRAW\ENJL1.TEX
%
%
\begin{figure}[htb]
\setlength{\unitlength}{1mm}
\begin{picture}(100.00,70.00)
\put(40.00,20.25){\vector(1,0){19.25}}
\put(58.50,20.25){\vector(1,0){20.00}}
\put(78.75,20.25){\line(1,0){14.50}}
\put(66.50,55.50){\circle{14.00}}
\put(66.50,27.50){\circle{14.00}}
\put(66.50,41.50){\circle{14.00}}
\put(66.50,20.00){\circle*{2.50}}
\put(65.50,48.25){\circle*{2.50}}
\put(66.50,34.50){\circle*{2.50}}
\put(65.00,61.50){X}
\put(50.00,10.00){Fig. 1a}
\end{picture}
\end{figure}
\begin{figure}[htb]
\unitlength 1cm
\begin{picture}(10,3)(-5,-0.5)
\thicklines
\put(-5,0){\vector(1,0){1}}
\put(-4,0){\line(1,0){1}}
\put(-2.5,-0.1){=}
\put(4,0){\line(1,0){1}}
\put(4,0.75){\circle{1.5}}
\put(4,0){\circle*{0.2}}
\thinlines
\put(3,0){\line(1,0){1}}
\put(2.5,-0.1){+}
\put(0,0){\vector(1,0){1}}
\put(1,0){\line(1,0){1}}

\put(-0.5,-2){Fig. 1b}

\end{picture}
\end{figure}
\vspace{1cm}
\begin{figure}[htb]
\setlength{\unitlength}{1mm}
\begin{picture}(140.00,50.00)
\thicklines
\put(97.50,35.00){\oval(15.00,10.00)}
\put(105.00,35.00){\circle*{2.00}}
\put(17.50,35.00){\oval(15.00,10.00)}
\put(25.00,35.00){\circle*{2.00}}
\put(32.50,35.00){\oval(15.00,10.00)}
\put(40.00,35.00){\circle*{2.00}}
\put(47.50,35.00){\oval(15.00,10.00)}
\put(55.00,35.00){\circle*{2.00}}
\put(62.50,35.00){\oval(15.00,10.00)}
\put(10.00,35.00){\circle*{2.00}}
\put(70.00,35.00){\circle*{2.00}}
\put(90.00,35.00){\circle*{2.00}}
\put(38.50,19.00){Fig. 2a}
\put(95.50,19.00){Fig. 2b}
\put(14.50,40.00){\vector(1,0){3.00}}
\put(29.50,40.00){\vector(1,0){3.50}}
\put(44.00,40.00){\vector(1,0){5.00}}
\put(60.50,40.00){\vector(1,0){3.00}}
\put(95.50,40.00){\vector(1,0){5.00}}
\put(99.00,30.00){\vector(-1,0){3.00}}
\put(64.00,30.00){\vector(-1,0){3.00}}
\put(49.50,30.00){\vector(-1,0){3.50}}
\put(34.00,30.00){\vector(-1,0){2.00}}
\put(18.00,30.00){\vector(-1,0){2.50}}
\end{picture}
\end{figure}
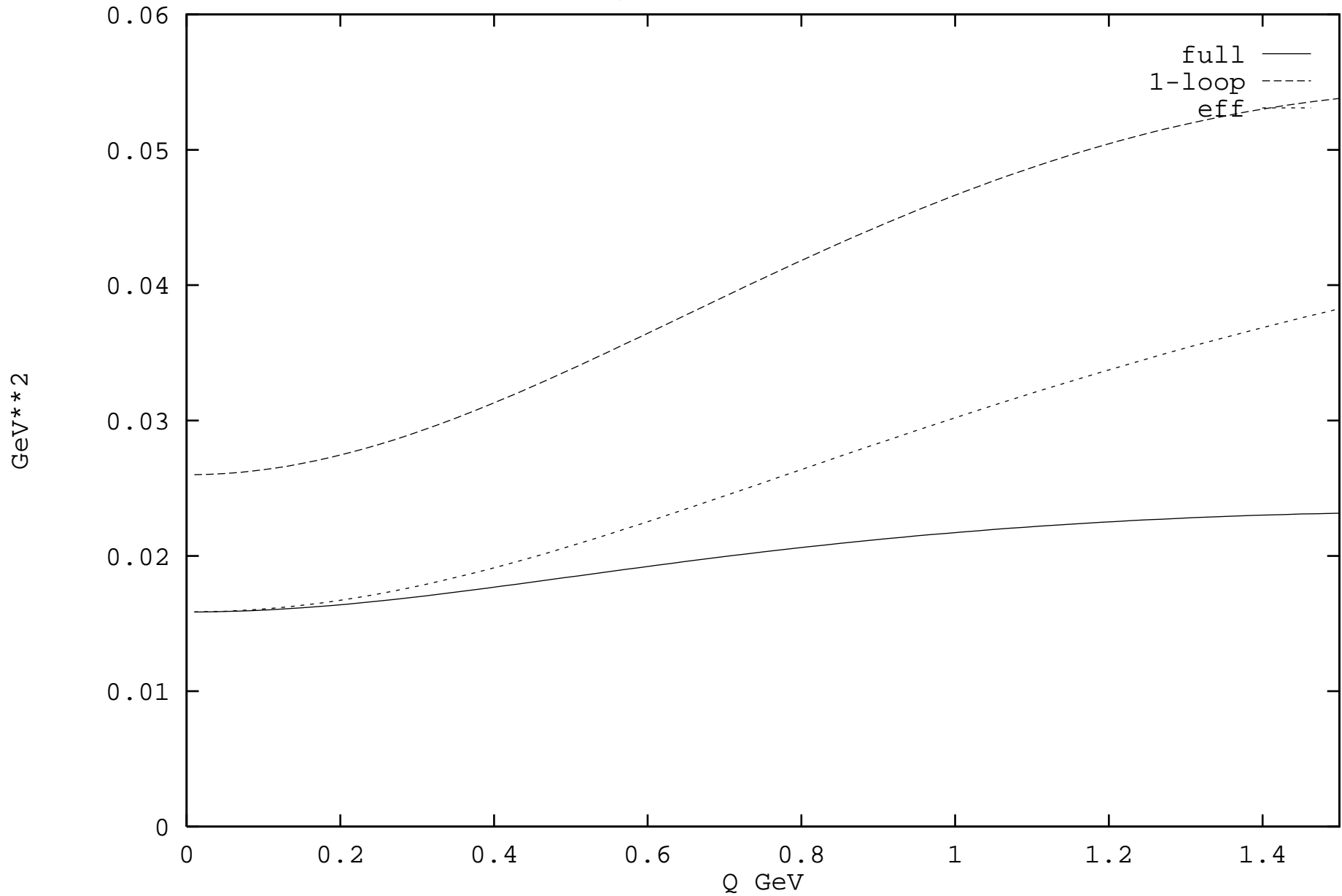
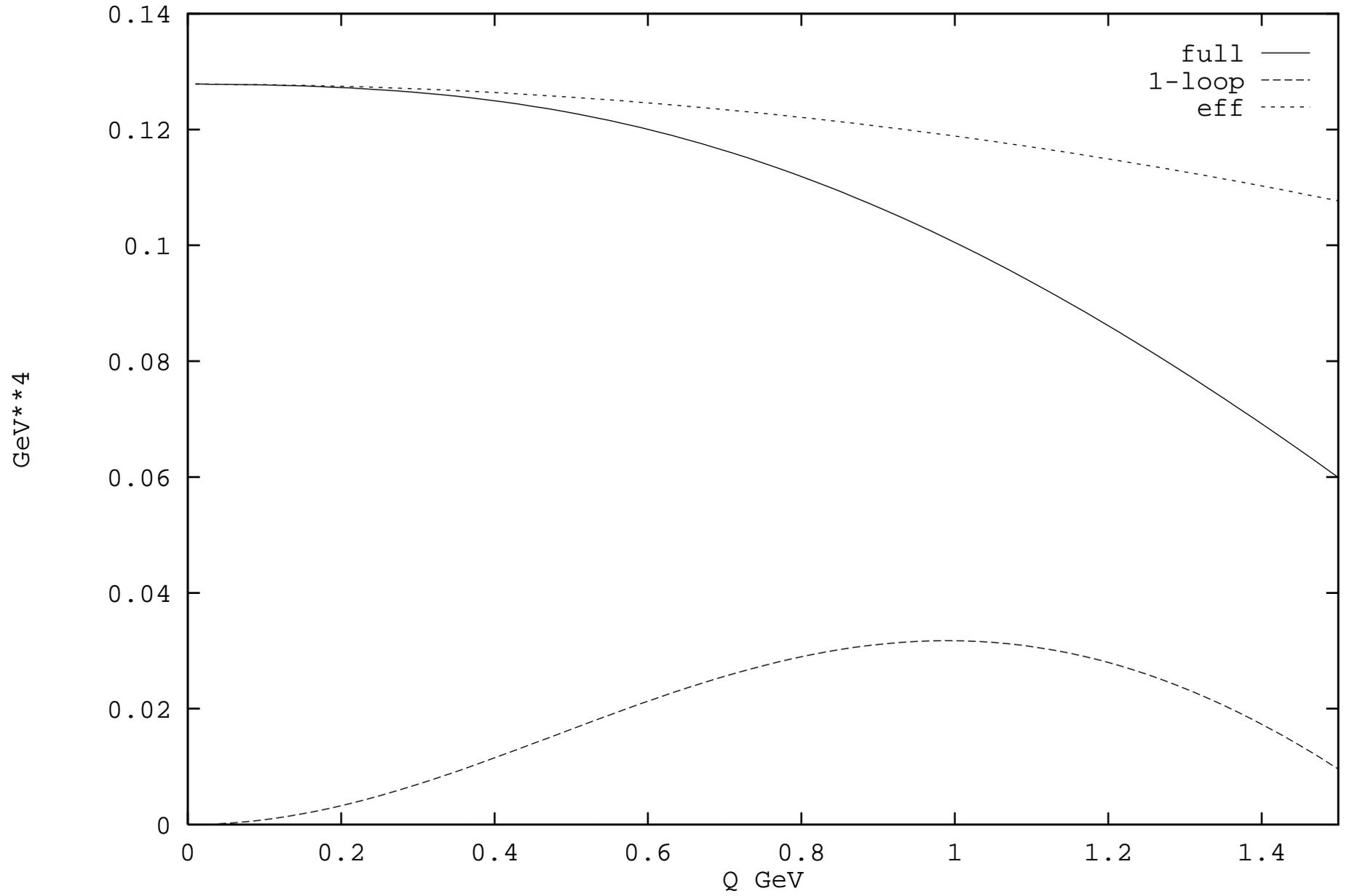
\end{document}